\documentclass[12pt]{article}
\usepackage{float}
\usepackage{subfig}
\usepackage[nottoc]{tocbibind}
\usepackage{authblk}
\usepackage{amsmath}
\usepackage{cite}
\usepackage{xcolor}
\usepackage{ulem}
\usepackage{amssymb}
\usepackage{graphicx}
\usepackage{comment}
\usepackage{epsfig}
\usepackage{epstopdf}
\usepackage{upgreek}

\usepackage{xcolor}
\usepackage[compat=1.1.0]{tikz-feynman}
\usepackage{tikzsymbols}
\usepackage{appendix}
\usepackage[colorlinks=true,urlcolor=black,linkcolor=black, citecolor=black]{hyperref}
\usepackage[a4paper, total={6.5in, 10.5in}]{geometry}
\usepackage{mathtools,slashed}
\usepackage{tikz}
\definecolor{deepmagenta}{rgb}{0.8, 0.0, 0.8}
\definecolor{mediumtealblue}{rgb}{0.0, 0.33, 0.71}
\definecolor{warmblack}{rgb}{0.0, 0.26, 0.26}
\definecolor{bostonuniversityred}{rgb}{0.8, 0.0, 0.0}
\definecolor{junglegreen}{rgb}{0.16, 0.67, 0.53}
\definecolor{lightcornflowerblue}{rgb}{0.6, 0.81, 0.93}

\title{\textbf{TeV scale leptogenesis with triplet fermion in connection to muon $g-2$ } }

\author[1]{\small{Simran Arora \thanks{009simranarora@gmail.com}}}

\affil[1,3]{\it{\small{Department of Physics and Astronomical Science, Central University of Himachal Pradesh, Dharamshala 176215,
INDIA.}}}

\author[2]{Devabrat Mahanta \thanks{devabrat@pragjyotishcollege.ac.in}}
\affil[2]{\it{Department of Physics, Pragjyotish College, Guwahati 781009, INDIA }}

\author[3]{B. C. Chauhan \thanks{bcawake@hpcu.ac.in}}

\begin{document}

\date{}

\maketitle

\begin{abstract}

    \noindent We propose an extension of the minimal scotogenic model with a triplet fermion and a singlet scalar.  An imposed $Z_{4}\times Z_{2}$ symmetry allows only diagonal Yukawa couplings among different generations of standard model (SM) leptons and right-handed singlet neutrinos. The Yukawa coupling of the triplet fermion with the inert doublet positively contributes to the muon anomalous magnetic moment.  The imposed $Z_{4}\times Z_{2}$ symmetry forbids the conventional leptogenesis from the lightest right-handed neutrino decay. A net lepton asymmetry can be generated in the muonic sector from $N_2$ and triplet fermion decay through resonant leptogenesis scenario. The Yukawa coupling of triplet plays significant role both in leptogenesis and in the anomalous magnetic moment of the muon. We show a viable parameter space for TeV scale leptogenesis while explaining the Fermi lab results. The inert scalar is the dark matter candidate in this model. The muon $(g-2)$ and dark matter both favor the same parameter space for mass of the dark matter and the triplet fermion. 

\noindent \textbf{Keywords:} $Z_4$ symmetry, muon ($g-2$), Fermion triplet, Leptogenesis, Scalar Dark Matter.
\end{abstract}

\newpage

\section{INTRODUCTION}

 Only $ \sim 5\%$ of our Universe is made up of baryonic matter. The baryonic matter has a large asymmetry between the particles and antiparticles. This asymmetry is termed as the baryon asymmetry of the Universe (BAU). The observed (BAU) is often quantified in terms of the baryon-photon ratio defined as \cite{Planck:2018vyg}

 \begin{equation}
     \eta_{B}= \dfrac{n_{B}-n_{\bar{B}}}{n_{\gamma}} \simeq 6.2 \times 10^{-10}.
 \end{equation}

\noindent While the measured value of the baryon to photon ratio from the cosmic microwave background (CMB) measurement agrees
with the big bang nucleosynthesis (BBN) estimate, the origin of baryon asymmetry has been a long-standing problem in particle physics and cosmology. To dynamically generate the required baryon asymmetry, three conditions were given by Sakharov \cite{Sakharov:1967dj} called "Sakharov conditions" as (1) baryon number violation, (2) $C$ and $CP$ violation and (3) out of equilibrium dynamics. All these requirements can be satisfied within the Standard Model (SM) with an expanding Universe. However, the required amount of asymmetry cannot be generated within the SM. Therefore, it demands new physics beyond the SM. Among different baryogenesis mechanisms, leptogenesis is one of the most promising mechanisms as it connects the high scale baryogenesis to the low-scale neutrino oscillation data (For a few reviews on leptogenesis see \cite{Buchmuller:2004nz, Davidson:2008bu, Pilaftsis:2009pk } and the references therein). In a conventional leptogenesis scenario, the decay of heavy right-handed neutrinos into SM lepton doublets and SM Higgs accounts for small lepton asymmetry of the Universe, which later gets converted into baryon asymmetry through $sphaleron$ process \cite{Khlebnikov:1988sr}. This mechanism is called Vanilla leptogenesis \cite{Minkowski:1977sc,Mohapatra:1980yp}. In the conventional method, with two right-handed neutrinos, the mass of right-handed neutrinos is high ($M_{1}\gtrsim 10^{9}$) GeV. Such high-mass right-handed neutrinos are difficult to probe in collider experiments. This motivates us to search for low-scale leptogenesis scenarios. In the minimal scotogenic model it is possible to have successful leptogenesis around the TeV scale \cite{Hugle:2018qbw} with three right-handed neutrinos. Although this is a minimal scenario to have low-scale leptogenesis, it is difficult to bring a connection with the observed excess in muon $(g-2)$ measurement by the Fermi Lab. Another possible way to lower the scale of leptogenesis is resonant leptogenesis (RL),\cite{Pilaftsis:2003gt, Pilaftsis:2005rv} where the self-energy effects enhance the lepton asymmetry due to tiny mass splitting between two fermion states. This is achieved when the mass difference between these states is comparable to their decay widths. It allows for successful leptogenesis at TeV scale. See \cite{Hambye:2001eu,Dev:2015cxa,Hugle:2018qbw,Borah:2021mri,Borah:2020ivi,Mahanta:2021plx,Singh:2023eye} and references therein for a few low-scale leptogenesis models. Here we consider a leptogenesis scenario where a triplet fermion resonantly enhances the $B-L_{\mu}$ asymmetry that is also responsible for generating muon $(g-2)$ excess.

\noindent Apart from the BAU, the existence of nonluminous dark matter (DM), giving a total contribution of around $26\%$ of the present Universe's energy density has been an unsolved puzzle in cosmology and particle physics. The DM energy density is expressed in the density parameter $\Omega_{\rm DM}$. The Planck 2018 \cite{Planck:2018vyg} data has reported the value for the DM density parameter  at $68\%$ CL

\begin{equation}
    \Omega_{\rm DM} h^{2}= 0.120\pm 0.001.
\end{equation}
 
\noindent Here $h=$ Hubble parameter/(100 Km $\rm s^{-1}$ $\rm Mpc^{-1}$). While none of the Standard Model particles satisfy the conditions to be a DM candidate, the SM also fails to adequately explain the origin of the baryon asymmetry. Another important problem the SM fails to explain is the existence of nonzero neutrino masses. It has led to a number of extensions of the SM. There have been multiple attempts where it is successfully addressed along with the BAU and DM production (For a detailed review of such ideas see \cite{DiBari:2019amk} and the references therein). 

 The recent measurement by the muon $(g-2)$ collaboration at Fermilab has reported a significant deviation in the anomalous magnetic moment $a_{\mu}=(g_{\mu}-2)/2$ of the muon from the standard model prediction \cite{Muong-2:2021ojo}. Upon combining Fermi lab results on muon ($g-2$) with the previous measurement from the Brookhaven National Laboratory shows a $4.2 \sigma$ observed excess of $\Delta a_{\mu} =251 (59) \times 10^{-11}$. This result has led to different BSM proposals (a comprehensive review can be found in \cite{Athron:2021iuf}). The authors of \cite{Borah:2021mri,Alvarez:2023dzz,Eijima:2023yiw} have shown few possible connections between the Fermi lab results on muon ($g-2$) and the BAU. On the other hand, the authors of \cite{Acuna:2021rbg, Arcadi:2021yyr, Bai:2021bau, Borah:2021khc,Borah:2022zim} had discussed few possible connections between DM and the muon $(g-2)$ excess. To the authors' knowledge, there have been no attempts to find a unified solution connecting both BAU and DM to the muon $(g-2)$ excess.

 In this work, we propose a minimal extension of the SM that can potentially address these key problems. We extend the minimal scotogenic model with a vector like fermion triplet by using a discrete $Z_{4}$ symmetry that only allows diagonal Yukawa matrices. The neutrino mass is generated through the scotogenic mechanism in the model. The decay of $N_2$ and fermion triplet with an almost degenerate mass around the TeV scale leads to net lepton asymmetry in the muon sector due to imposed $Z_{4} \times Z_{2}$ symmetry. The fermion triplet also has its role in explaining muon ($g-2$) through its coupling with inert doublet which is the dark matter candidate in the model.

 The rest of the paper is organized as follows: In Sec. \ref{sec:model}, we briefly discuss our basic framework and the corresponding Lagrangian. Sec. \ref{sec:nu_mass} provides a detailed description of the neutrino mass mechanism through the scotogenic model. In Sec. \ref{sec:muon}, we introduce the muon ($g-2$) and the results for the proposed model, in Sec. \ref{sec:DM}, we provide the analysis of inert dark matter in the model. In Sec. \ref{sec:lepto}, low-scale leptogenesis by out of equilibrium decay of the right-handed neutrino $N_2$ assisted by fermion triplet is explained. Sec. \ref{sec:results} presents the detailed summary of the results from the model. In Section \ref{sec:conclusion}, conclude the work.

\section{THE MODEL}

\label{sec:model}

The scotogenic neutrino mass generation proposed in \cite{Ma:2006km} has been a very promising minimal extension of the SM explaining the origin of nonzero neutrino mass. The scotogenic model not only provides an elegant way of DM production but also can generate the observed baryon asymmetry at relatively low scales. The authors \cite{Hugle:2018qbw} have shown the possibility of a TeV scale leptogenesis in the minimal scotogenic model. A few different leptogenesis scenarios in the context of the scotogenic model can be found in \cite{Borah:2018rca,Mahanta:2019gfe,Borah:2020ivi}. In this paper, we propose extending the scotogenic model with a triplet fermion. The particle content of the model is shown in table (\ref{tab1}). The SM is extended by three copies of right-handed neutrinos ($N_{i}$), a vectorlike triplet fermion ($\psi$), a scalar singlet ($S$) and an inert doublet ($\eta$). All the fields are charged under an imposed $Z_{4}\times Z_{2}$ symmetry as shown in the Table (\ref{tab1}).

\begin{table}[h] \label{Tab1}
\centering
 \begin{tabular}{c c c c c c c c} 
 \hline\hline

 Symmetry Group & $L_{e}$, $L_{\mu }$, $L_{\tau }$ & $e_{R}$, $\mu_{R}$, $\tau_{R}$ &$N_{1}$, $N_{2}$, $N_{3}$ & $\psi$ & $H$ & $S$ & $\eta$ \\ [0.5ex] 
 \hline
 $SU(2)_{L}$ $\times$ $U(1)_{Y}$ & (2, -1/2) & (1, -1) & (1, 0)&(3, -1) & (2, 1/2) & (1, 0)&(2, 1/2) \\ 
$ Z_{4}$ &($1, i, -i$) & (1, $i, -i$) & (1, $i, -i$)&$i$ &1&$-i$&1 \\
$Z_{2}$ & + & + & -- & -- & +&+& -- \\ [1ex] 
 \hline
 \end{tabular}
  \caption{\small{The field content and respective charge assignments of the model under $SU(2)_{L}\times U(1)_{Y}\times Z_{4}\times Z_{2}$.}}
      \label{tab1}
\end{table}

\noindent The relevant terms in the Yukawa Lagrangian are

    \begin{eqnarray} \label{Ly}
    \nonumber
-\mathcal{L} &\supseteq&  \frac{M_{11}}{2} N_{1} N_{1} + M_{23} N_{2} N_{3} + y_{\eta 1} \overline{L}_{e} \tilde{\eta} N_{1} + y_{\eta 2} \overline{L}_{\mu} \tilde{\eta} N_{2 } + y_{\eta 3} \overline{L}_{\tau} \tilde{\eta} N_{3} \nonumber \\ &&
    +\:  y_{12} S  N_{1} N_{2} + y_{13} S^{*} N_{1} N_{3} +  y_{\psi} 
 \overline{L_{\mu}}C {\psi}^{\dagger}\tilde{\eta} + m_{\psi} \overline{\psi} (\psi)^{c} \nonumber \\ &&
    +\: y_{ e} \overline{L}_{e} H e_{R} + y_{\mu} \overline{L}_{\mu} H \mu_{R} + y_{\tau} \overline{L}_{\tau} H \tau_{R} + H.c.,
   \end{eqnarray}
with $\tilde{\eta}=i \sigma_{2}\eta^{*}$ and the scalar potential $V(H,S,\eta)$ is given by 
  
   \begin{eqnarray}
    \nonumber
    V(H,S,\eta) & = & -\mu_{H}^{2}(H^{\dagger} H)+\lambda_{1}(H^{\dagger} H)^{2}-\mu_{S}^{2}(S^{\dagger}S)
     +\lambda_{S}(S^{\dagger
}S)^{2}+\lambda_{HS}(H^{\dagger} H)(S^{\dagger} S) \nonumber \\ && +\:\mu_{\eta}^{2}(\eta^{\dagger}\eta) 
+ \lambda_{2}(\eta^{\dagger}\eta)^{2} + \lambda_{3}(\eta^{\dagger}\eta)(H^{\dagger} H)+\lambda_{4}(\eta^{\dagger} H)(H^{\dagger} \eta)\nonumber\\
     && \: +\frac{\lambda_{5}}{2}[(H^{\dagger} \eta)^{2}+(\eta^\dagger H)^{2}]+
     \lambda_{\eta S}(\eta^{\dagger} \eta)(S^{\dagger} S) + H.c..
\end{eqnarray}

\noindent The SM gauge symmetry is broken by the neutral component of Higgs doublet $H$ while the $Z_{4}$ symmetry is spontaneously broken by the nonzero vacuum expectation value ($\rm VEV$) of scalar singlet $S$. Also, we assume $\mu_{\eta}^2$ $>$ 0  so that $\eta$ do not acquire any $\rm VEV$. Hence the masses of singlet scalar and Higgs are given as 

\begin{equation}
    m_{S,H}^2 = \lambda_{1}v^2 + \lambda_{S}v_{S}^{2} \pm (\lambda_{1}v^{2}+\lambda_{S}v_{S}^{2})\sqrt{1+r^2},
\end{equation}
where $r=\frac{\lambda_{HS}v v_{S}}{\lambda_{1}v^2 - \lambda_{S}v_{S}^2}.$ The masses of neutral and charged components of inert doublet are written as

\begin{eqnarray}
  m_{\eta^{\pm}}^2 & = & \mu_{\eta}^{2} + \frac{1}{2}\lambda_{3}v^{2} + \frac{1}{2}\lambda_{\eta S}v_{S}^{2},  \\
  m_{\eta_R}^{2} & = & \mu_{\eta}^{2} + \frac{1}{2}\lambda_{3}v^{2} + \frac{1}{2}(\lambda_{4}+\lambda_{5})v^2 + \frac{1}{2} \lambda_{\eta S}v_{S}^{2} , \\
   m_{\eta_I}^{2} & = & \mu_{\eta}^{2} + \frac{1}{2}\lambda_{3}v^{2} + \frac{1}{2}(\lambda_{4}-\lambda_{5})v^2 + \frac{1}{2} \lambda_{\eta S}v_{S}^{2} .
\end{eqnarray}

\noindent Here it is important to note that $m_{\eta_{R}}^{2}-m_{\eta_{I}}^{2}=\lambda_{5}v^{2}$. Using Eq. (\ref{Ly}), the charged lepton mass matrix $M_{l}$,  Dirac Yukawa matrix $y_{D}$ and right-handed neutrino mass matrix $M_{R}$ are given by

\begin{eqnarray*}
M_{l} = \frac{1}{\sqrt{2}}\begin{pmatrix}
   y_{e}v && 0 && 0 \\
   0 &&  y_{\mu}v && 0 \\
   0 && 0 &&  y_{\tau}v
   \end{pmatrix},
\end{eqnarray*}

\begin{eqnarray}
  y_{D} = 
   \begin{pmatrix}
   y_{\eta 1} && 0 && 0 \\
   0 &&  y_{\eta 2} && 0 \\
   0 && 0 &&  y_{\eta 3}
   \end{pmatrix}  , M_{R} =
    \begin{pmatrix}
   M_{11} && y_{12}v_{S}/\sqrt{2} && y_{13}v_{S}/\sqrt{2} \\
   y_{12}v_{S}/\sqrt{2} &&  0 && M_{23}e^{i\delta} \\
   y_{13}v_{S}/\sqrt{2} && M_{23}e^{i\delta} && 0
   \end{pmatrix}.
\end{eqnarray}
Here $v/\sqrt{2}$ and $v_{S}/\sqrt{2}$ are $vevs$ of the Higgs field $H$ and scalar field $S$, respectively, and $\delta$ is the phase remaining after redefinition of the fields.

\subsection{Neutrino mass}

\label{sec:nu_mass}

In this model, neutrinos get mass by the scotogenic mechanism \cite{Ma:2006km}. The light neutrino mass matrix is given by 
 
 \begin{equation}\label{scot}
     M_{ij}^{\nu} = \sum_{k=1}^{3}{}\frac{d_{ik}d_{jk}M_{k}}{32\pi^{2}}\left[L_{k}(m_{\eta_{R}}^{2}) - L_{k}(m_{\eta_{I}}^{2})  \right],
 \end{equation}
where 
  \begin{equation}
  L_{k}(m^{2}) = \frac{m^{2}}{m^{2}-M_{k}^{2}} ln \frac{m^2}{M_{k}^{2}}.     
  \end{equation}
 
\noindent  Here $M_{k}$ is the mass of $k^{th}$ right-handed neutrino and $m_{\eta_{R},\eta_{I}}$ are the masses of real and imaginary parts of inert doublet $\eta$ and indices $i$, $j = 1, 2, 3$ run over three neutrino generations. Here $d_{ik}=y_{\eta i} C_{ik}$, where $C_{ik}$ are the elements of the matrix that diagonalize $M_{R}$. We use the Cassas-Ibarra (CI) \cite{Casas:2001sr} parametrization for radiative seesaw mechanism \cite{Toma:2013zsa} through which the Yukawa coupling satisfy neutrino oscillation data. The parametrization is given by 
  \begin{equation}
      d_{i k} = (\text{U} D_{\nu}^{1/2} R^{\dagger} \Lambda ^{1/2})_{i k}.
      \label{Eq:CI}
  \end{equation}
  Here R is a complex orthogonal matrix, $D_{\nu}=diag(m_1,m_2,m_3)$ is diagonal light neutrino mass matrix. The entries of the diagonal matrix $\Lambda$ is given by 
  \begin{equation}
      \Lambda_{k} = \frac{2 \pi^{2}}{\lambda_{5}}\zeta_{k}\frac{2 M_{k}}{v^2},
  \end{equation}
  where \begin{equation}
     \zeta_{k} = \bigg(\frac{M_{k}^{2}}{8 (m_{\eta_{R}^{2}}-m_{\eta_{I}^{2}})}\left[L_{k}(m_{\eta_{R}}^{2})- L_{k}(m_{\eta_{I}}^{2})\right]\bigg).
  \end{equation}

\subsection{Muon $(g-2)$}\label{sec:muon}

\begin{figure}[htb!]
    \centering
    \includegraphics[scale=0.4]{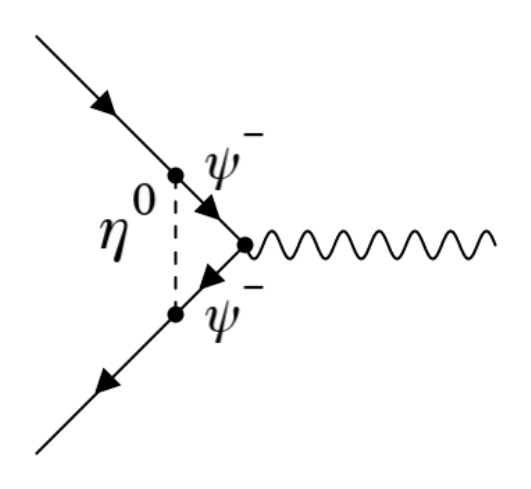}
    \includegraphics[scale=0.4]{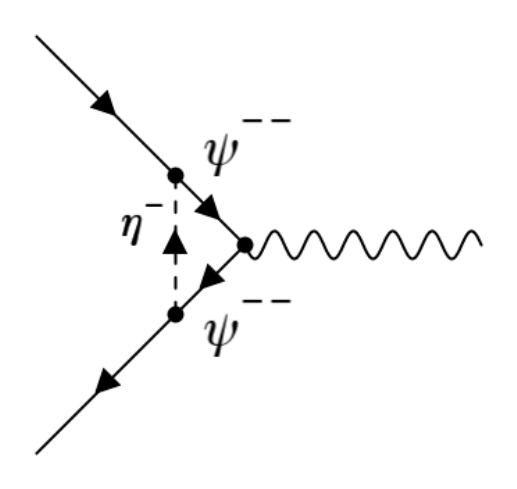}
    \includegraphics[scale=0.4]{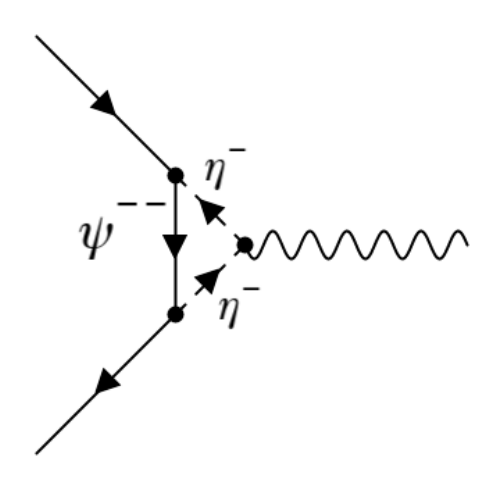}
    \caption{\small{Feynman diagrams showing the contribution of fermion triplet to muon ($g-2$).}}
    \label{f1}
\end{figure}

The fermion triplet $\psi$ explains muon $(g-2)$ by its coupling with SM muon and inert
doublet $\eta$. The Feynman diagrams for the same are given in Fig. \ref{f1}. The additional contribution to muon $(g-2)$ is given as \cite{Freitas:2014pua}  

\begin{eqnarray} \label{au2}
	\Delta a_{\mu} = \frac{m_{\mu}^2 y_{\psi}^2}{32 \pi^2 m_{\eta}^2}[5 f_{1}(m_{\psi}^2/m_{\eta}^2) - 2 f_{2}(m_{\psi}^2/m_{\eta}^2)],
\end{eqnarray}

\noindent where  
$m_{\mu}$, $m_{\psi}$, $m_{\eta}$ are the masses of muon, fermion triplet $\psi$, inert scalar doublet $\eta$ respectively. Here we keep the masses of the neutral scalars ($\eta_{R}, \eta_{I}$) and the charged scalars ($\eta^{\pm}$) nearly the same and are represented by $m_{\eta}$. The loop functions are given by 
\begin{eqnarray*}
  f_{1}(x) = \frac{1}{6(x-1)^4}[x^3 - 6x^2 + 3x + 2 + 6x \ln x],
\end{eqnarray*}
\begin{eqnarray}
   f_{2}(x) = \frac{1}{6(x-1)^4}[-2x^3 - 3x^2 + 6x - 1 + 6x^2 \ln x].
\end{eqnarray}

\begin{figure}[h]
\centering
\includegraphics[scale=0.45]{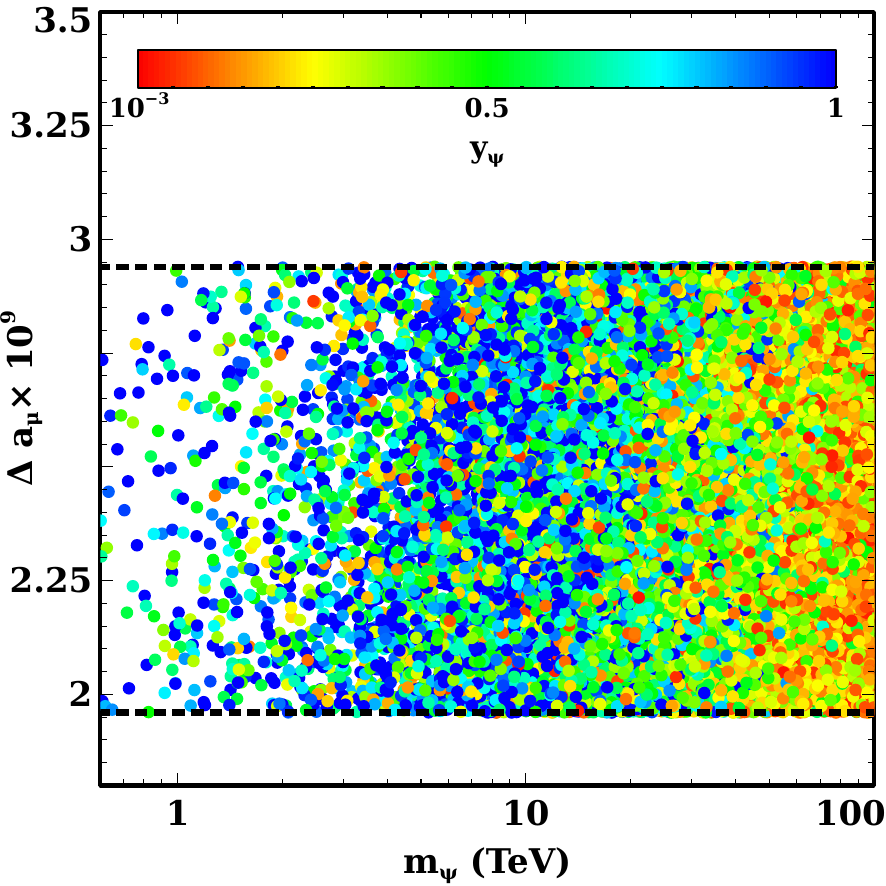}
\includegraphics[scale=0.45]{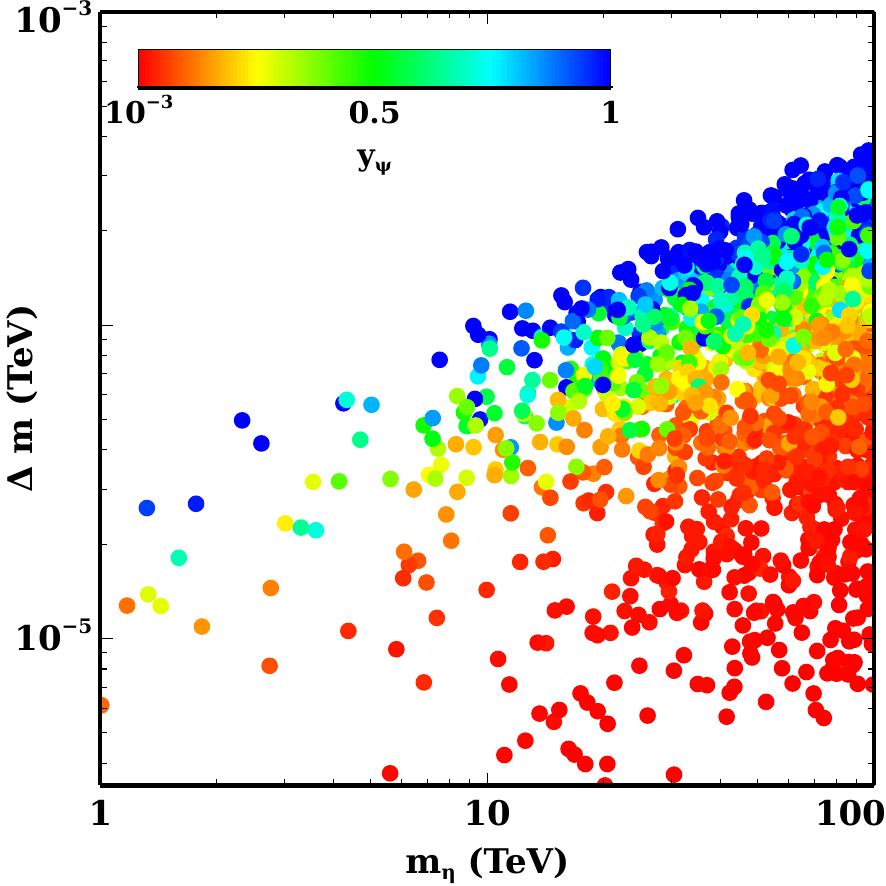}
  \caption{\small{Plots showing the data points allowed by the Fermi lab's result on muon $(g-2)$. Right panel shows the parameter space in $m_{\eta}-\Delta m$ plane with $y_{\psi}$ as the color bar   
  and the left panel shows the allowed data points with $m_{\psi}$ and $\Delta a_{\mu} =251 (59) \times 10^{-11}$. Here $m_{\psi}$ is varied randomly from 0.75 TeV to 100 TeV and $m_{\eta} = m_{\psi} - \Delta m$, with $\Delta m$ varied from $10^{-6}$ TeV to $10^{-3}$ TeV. }  }
  \label{fig:muon}
\end{figure}

\noindent Here, $x$ is a dimensionless parameter given by $x=m_{\psi}^2/m_{\eta}^2$. In Fig. (\ref{fig:muon}) we show the data points that satisfy the Fermi lab result on $\Delta a_{\mu}$. In the left panel plot, we show the data 
 points within the allowed range of $\Delta a_{\mu}$ from the Fermi Lab result on muon $g-2$ with $m_{\psi}$ in the x-axis and $y_{\psi}$ as color bar. In the right panel plot, we show the allowed parameter space from the Fermi Lab result on muon $(g-2)$ in $m_{\eta}-\Delta m$ plane, $\Delta m=m_{\psi}-m_{\eta}$ being the mass difference between the triplet fermion $\psi$ and the doublet scalar $\eta$. The Yukawa coupling $y_{\psi}$ is shown by the color bar. From the right panel plot of Fig. (\ref{fig:muon}) it can be seen that to satisfy the Fermi lab results on muon $(g-2)$ excess one need $m_{\eta} \simeq m_{\psi}$. With $m_{\eta} \simeq m_{\psi}$ the contribution to muon $(g-2)$ coming from the diagrams shown in Fig. \ref{f1} enhances. To produce the observed excess of muon $(g-2)$ we require the  $m_{\eta} \simeq m_{\psi}$ even with the maximum possible value of $y_{\psi}=4\pi$. Here we take the Yukawa coupling in the range $0.001\leq y_{\psi} \leq 1$. From the left panel plot in Fig. \ref{fig:muon}, it can be seen that there exists a correlation between $m_{\psi}$ and $y_{\psi}$. With the increase in $m_{\psi}$ we require lower values of  $y_{\psi}$ to generate the observed Muon $g-2$ excess. Similarly from the right panel plot in Fig. \ref{fig:muon} one can notice a positive correlation between the $\Delta m$ and $y_{\eta}$. With the increase in $\Delta m$, the required value of $y_{\psi}$ increases. Also it is seen that there is a negative correlation between $m_{\eta}$ and $y_{\psi}$.

\subsection{Dark matter}

\label{sec:DM}

In the minimal scotogenic model, scalar and fermionic DM scenarios are possible depending on the lightest $Z_{2}$ odd state. The lightest of the neutral components of the inert doublet $\eta$ serves to be a weakly interacting massive particle (WIMP) DM. This scenario is extensively studied in the literature \cite{Dolle:2009fn, LopezHonorez:2010eeh, Mahanta:2019gfe}. If the inert doublet scalar becomes heavier than the lightest right-handed neutrino, the latter becomes a DM candidate. Depending on the size of Yukawa couplings, the fermionic DM can be produced thermally or nonthermally \cite{Mahanta:2019gfe}. Here, we take the scalar DM scenario, where the real part of the neutral component of the inert doublet is the DM particle. The model is implemented in FEYNRULES \cite{Alloul:2013bka,Christensen:2008py} to compute all the relevant vertices. To solve the Boltzmann equations and to calculate the relic density of dark matter, micrOMEGAs is used \cite{Belanger:2010pz,Belanger:2013ywg}. Apart from the annihilation and coannihilation of inert doublet DM, here we also have a few additional coannihilation of the DM with the triplet fermion $\psi$. The coannihilation processes are shown in Fig. \ref{fig:Feynman}. The relevant cross sections for the coannihilation processes are shown in Appendix \ref{appen1}.
\begin{figure}[t]
    \centering
\begin{center}
\begin{tikzpicture}[scale=0.7]
  \begin{feynman}
    \vertex (i1) at (-3, 1.5) {\(\boldsymbol{\psi}\)};
    \vertex (i2) at (-3, -1.5) {\(\boldsymbol{\eta}\)};
    
    \vertex (v1) at (0, 1.5);  
    \vertex (v2) at (0, -1.5); 
    \vertex (c) at (0, 1.5);    
    
    \vertex (o1) at (3, 1.5) {\(\boldsymbol{L_{\mu}}\)};
    \vertex (o2) at (3, -1.5) {\(\boldsymbol{h}\)};
    
    \diagram* {
      (i1) -- [plain, very thick] (v1) -- [plain, very thick] (c),
      (i2) -- [scalar, very thick] (v2) -- [scalar, very thick,edge label'=\(\boldsymbol{\eta}\)] (c),
      (c) -- [scalar,very thick,edge label'=\(\boldsymbol{\eta}\)] (c),
      (c) -- [scalar,very thick] (v1) -- [fermion, very thick] (o1),
      (c) -- [scalar, very thick] (v2) -- [scalar, very thick] (o2),
    };
  \end{feynman}
\end{tikzpicture}
\end{center}
\begin{center}
\begin{tikzpicture}[scale=0.7]
  \begin{feynman}
    \vertex (i1) at (-2.5, 1.5) {\(\boldsymbol{\psi}\)};
    \vertex (i2) at (-2.5, -1.5) {\(\boldsymbol{\eta}\)};
    
    \vertex (v1) at (0, 1.5);  
    \vertex (v2) at (0, -1.5); 
    \vertex (c) at (0, 1.5);    
    
    \vertex (o1) at (3, 1.5) {\(\boldsymbol{W^{\pm}/Z/\gamma}\)};
    \vertex (o2) at (3, -1.5) {\(\boldsymbol{L_{\mu}}\)};
    
    \diagram* {
      (i1) -- [plain, very thick] (v1) -- [plain, very thick] (c),
      (i2) -- [scalar, very thick] (v2) -- [plain, very thick,edge label'=\(\boldsymbol{\psi}\)] (c),
      (c) -- [scalar,very thick,edge label'=\(\boldsymbol{\eta}\)] (c),
      (c) -- [scalar,very thick] (v1) -- [red,photon, very thick] (o1),
      (c) -- [scalar, very thick] (v2) -- [fermion, very thick] (o2),
    };
  \end{feynman}
\end{tikzpicture}
\quad
\begin{tikzpicture}[scale=0.7]
  \begin{feynman}
    \vertex (i1) at (-2, 1.5) {\(\boldsymbol{\psi}\)};
    \vertex (i2) at (-2, -1.5) {\(\boldsymbol{\eta}\)};
    
    \vertex (v1) at (0, 0);
    \vertex (v2) at (3, 0);
    
    \vertex (o1) at (5.5, 1.5) {\(\boldsymbol{W^{\pm}/Z/\gamma}\)};
    \vertex (o2) at (5, -1.5) {\(\boldsymbol{L_{\mu}}\)};
    
    \diagram* {
      (i1) -- [plain, very thick] (v1) -- [fermion, very thick, edge label=\(\boldsymbol{L_{\mu}}\)] (v2) -- [red,photon, very thick] (o1),
      (i2) -- [scalar, very thick] (v1),
      (v2) -- [fermion, very thick] (o2),
    };
  \end{feynman}
\end{tikzpicture}
\end{center}

\begin{center}
\begin{tikzpicture}[scale=0.7]
  \begin{feynman}
    \vertex (i1) at (-3, 1.5) {\(\boldsymbol{\psi}\)};
    \vertex (i2) at (-3, -1.5) {\(\boldsymbol{\eta}\)};
    
    \vertex (v1) at (0, 1.5);  
    \vertex (v2) at (0, -1.5); 
    \vertex (c) at (0, 1.5);    
    
    \vertex (o1) at (3, 1.5) {\(\boldsymbol{N_{2}}\)};
    \vertex (o2) at (3, -1.5) {\(\boldsymbol{\eta}\)};
    
    \diagram* {
      (i1) -- [plain, very thick] (v1) -- [fermion, very thick] (c),
      (i2) -- [scalar, very thick] (v2) -- [fermion, very thick,edge label'=\(\boldsymbol{L_{\mu}}\)] (c),
      (c) -- [fermion,very thick,edge label'=\(\boldsymbol{L_{\mu}}\)] (c),
      (c) -- [fermion,very thick] (v1) -- [plain, very thick] (o1),
      (c) -- [scalar, very thick] (v2) -- [scalar, very thick] (o2),
    };
  \end{feynman}
\end{tikzpicture}
\quad
\begin{tikzpicture}[scale=0.7]
  \begin{feynman}
    \vertex (i1) at (-2.5, 1.5) {\(\boldsymbol{\psi}\)};
    \vertex (i2) at (-2.5, -1.5) {\(\boldsymbol{\eta}\)};
    
    \vertex (v1) at (0, 0);  
    \vertex (v2) at (3, 0); 

    \vertex (o1) at (5.5, 1.5) {\(\boldsymbol{N_{2}}\)};
    \vertex (o2) at (5.5, -1.5) {\(\boldsymbol{\eta}\)};
    
    \diagram* {
      (i1) -- [plain, very thick] (v1) -- [fermion, very thick, edge label=\(\boldsymbol{L_{\mu}}\)] (v2) -- [plain, very thick] (o1),
      (i2) -- [scalar, very thick] (v1),
      (v2) -- [scalar, very thick] (o2),
    };
    
  \end{feynman}
\end{tikzpicture}
\end{center}
    \caption{\small{Coannihilation channels of dark matter with triplet fermion $\psi$.}}
    \label{fig:Feynman}
\end{figure}

\begin{figure}[htb!]

\centering
\includegraphics[scale=0.45]{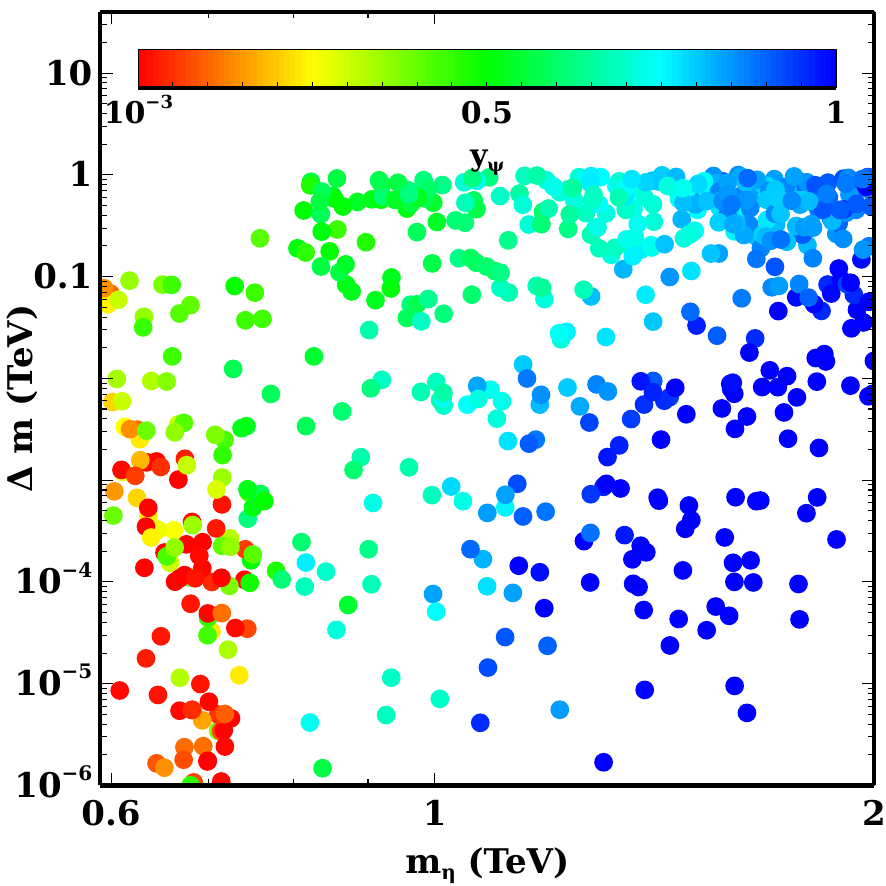}
\includegraphics[scale=0.45]{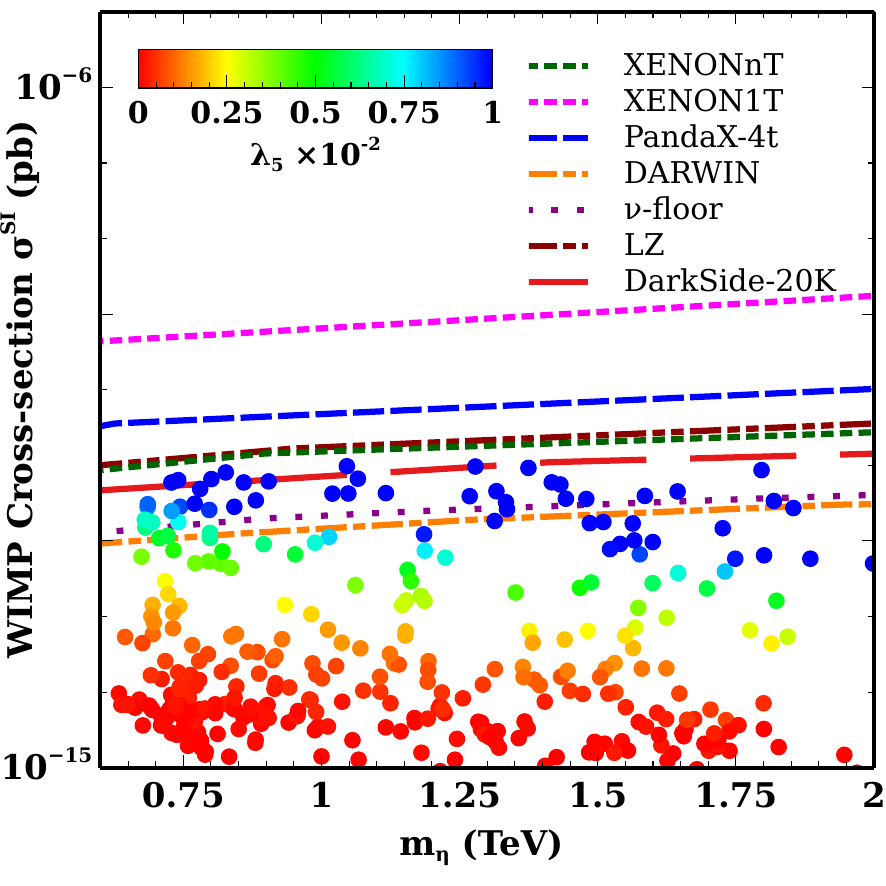}
\includegraphics[scale=0.45]{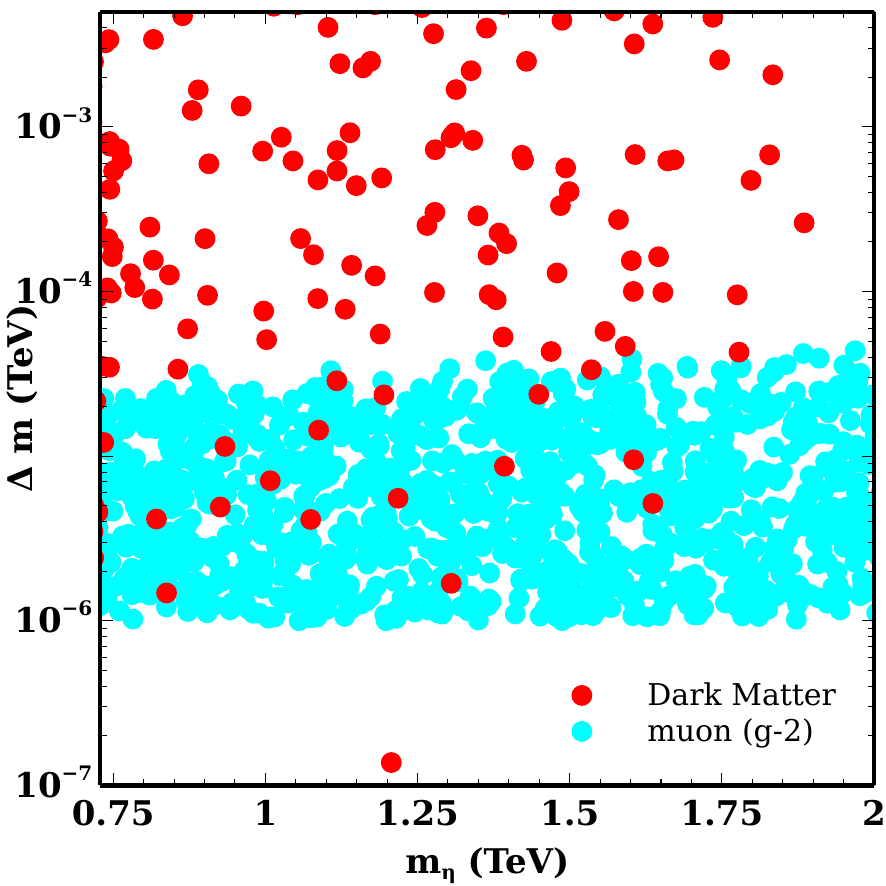}

\caption{\small{Scan plot showing the parameter space in $m_{\eta}-\Delta m$ plane with $y_{\psi}$ as the color bar that satisfy the DM relic ( upper left panel), DM spin independent direct detection cross section ( upper right panel) and common parameter space in $m_{\eta}-\Delta m$ plane for muon ($g-2$) and dark matter (lower panel). Here $m_{\psi}$ is varied randomly from 0.6 TeV to 2 TeV and $m_{\eta} = m_{\psi} - \Delta m$, with $\Delta m$ varied from $10^{-6}$ TeV to 1 TeV. The couplings $\lambda_3, \lambda_4$ and $\lambda_5$ are varied in the range $(10^{-5} -1)$.} }
\label{fig:DM}
 
\end{figure}

\noindent To satisfy the correct relic, the parameter space is constrained in $m_{\psi}-y_{\psi}$ plane which is relevant for leptogenesis and muon $(g-2)$ as well. On the top left panel plot of Fig.\ref{fig:DM}, we show the parameter space allowed by the correct DM relic in $m_{\eta}-\Delta m$ plane. Since the mass required for the doublet scalar to satisfy the Fermi lab result on muon $(g-2)$ lies in the TeV range, we consider $m_{\eta} \ge 600$ GeV for the DM analysis. In this region of parameter space, apart from annihilation and coannihilation among the inert doublet components, we have additional coannihilation channels of the inert doublet DM with $\psi$. The small mass difference required by the muon $(g-2)$ excess makes the coannihilation between DM and $\psi$ significant. On the top right panel plot of Fig.\ref{fig:DM} we plot the spin independent direct detection cross sections with DM mass for the relic satisfied points. We show all the recent DM direct search limits from the XENON1T \cite{XENON:2020rca}, PandaX-4t \cite{PandaX-4T:2021bab}, XENONnT \cite{XENON:2023cxc}, LZ \cite{LZ:2022lsv}, and the proposed sensitivity limit from the upcoming experiments Darwin \cite{DARWIN:2016hyl} and DarkSide-20K \cite{DarkSide-20k:2017zyg}. In the lower panel plot of Fig.\ref{fig:DM} we show the allowed parameter spaces required to satisfy the correct DM relic and the observed muon $(g-2)$ by the Fermi Lab. The cyan points generate the observed muon $g-2$ by the Fermi Lab while the red points satisfy the correct DM relic.

\subsection{Leptogenesis}
\label{sec:lepto}

In this model, a net lepton asymmetry cannot be generated from the decay of the lightest right handed neutrino ($N_{1}$) as there are no one loop diagram to generate the required $CP$ asymmetry. A net lepton asymmetry can be generated from the decay of the right handed neutrino $N_{2}$ and the neutral component of $\psi$. In the case of hierarchical mass spectrum ($m_{\psi}\ll M_{2}$ or $M_{2}\ll m_{\psi}$) the lepton asymmetry is mainly generated by the decay of the lightest fermion. The asymmetry produced from the decay of the heavier fermion is washed out by the inverse decay of the lighter one. If we take the hierarchy $m_{\psi}< M_{2}$ a lepton asymmetry is generated in muon/muonic neutrino from the decay of $\psi$. A few earlier works on leptogenesis from a triplet fermion can be found in \cite{AristizabalSierra:2010mv,Das:2022qyc,Vatsyayan:2022rth,Singh:2023eye,Mahapatra:2023dbr,Borah:2024wos}.
Being triplet, $\psi$ has strong gauge scatterings that keep it in equilibrium up to a small temperature. It reduces the asymmetry production. This result in a high scale leptogenesis $(m_{\psi}\gtrsim 10^{10}$ GeV). Similarly if we take ($M_{2}< m_{\psi}$) a lepton asymmetry can be generated in the muon/muonic neutrino from the decay of $N_{2}$. In a scotogenic neutrino mass generation the Yukawa coupling of $N_{2}$ is always large enough resulting in a strong washout of asymmetry from the inverse decays. This region of parameter space is known as a strong washout region $(\Gamma_{2}/H(T=M_{2})\gtrsim 4)$. In a strong washout region, it is impossible to have leptogenesis around the TeV scale with hierarchical mass spectrum \cite{Mahanta:2019gfe}. To be consistent with muon ($g-2$) and dark matter parameter space the only way to have TeV scale leptogenesis is to consider a resonant leptogenesis \cite{Pilaftsis:2003gt}. The resonant condition would be satisfied if $M_{2} -m_{\psi}  \simeq \Gamma_{2,\psi}/2$, $\Gamma_{2,\psi}$ being the decay width of $N_{2}$ and $\psi$ respectively. In a resonant leptogenesis scenario since the masses of $N_{2}$ and $\psi$ are similar, both can contribute to the production of a lepton asymmetry. The relevant tree level and self-energy diagram for generating $CP$ asymmetry in $N_{2}$ and $\psi$ decay are shown in Fig \ref{fig:lepto_loop}. 
The $CP$ asymmetry parameters generated from the interference of tree level and one loop self-energy diagrams are given in Eq. \ref{eq: CPasymmetry}. 
 
\begin{figure}[htb!]
\begin{center}
\begin{tikzpicture}[scale=0.6]
      \begin{feynman}
        \vertex (i1) at (-3, 0) {\(\boldsymbol{N_2}\)};
        \vertex (v1) at (0, 0);
        \vertex (o1) at (2, 1.5) {\(\boldsymbol{L_{\mu}}\)};
        \vertex (o2) at (2, -1.5) {\(\boldsymbol{\eta}\)};
        
        \diagram* {
          (i1) -- [plain,very thick] (v1),
          (v1) -- [fermion, very thick] (o1),
          (v1) -- [scalar,very thick] (o2),
        };
      \end{feynman}
    \end{tikzpicture}
    \hspace{1cm} 
    \begin{tikzpicture}[scale=0.6]
      \begin{feynman}
        \vertex (i1) at (-3, 0) {\(\boldsymbol{N_2}\)};
        \vertex (v1) at (0, 0);
        \vertex (loop1) at (2, 0); 
        \vertex (loop2) at (2, 4); 
        \vertex (v2) at (2, 0);
        \vertex (v3) at (4,0);
        \vertex (o1) at (7, 1.2) {\(\boldsymbol{L_{\mu}}\)};
        \vertex (o2) at (7, -1.2) {\(\boldsymbol{\eta}\)};
        
        \diagram* {
          (i1) -- [plain,very thick] (v1) -- [white,plain] (v2)  -- [plain,very thick,edge label=\(\boldsymbol{{\psi}}\)] (v3)-- [fermion,very thick] (o1),
          (v3) -- [scalar,very thick] (o2),
          (v1) -- [scalar,very thick, half left, edge label=\(\boldsymbol{{\eta}}\)] (loop1) -- [fermion, very thick,half left, edge label=\(\boldsymbol{L_{\mu}}\)] (v1) 
        };
      \end{feynman}
    \end{tikzpicture}
    \end{center}
\quad
    \begin{center}
    \begin{tikzpicture}[scale=0.6]
      \begin{feynman}
        \vertex (i1) at (-3, 0) {\(\boldsymbol{\psi}\)};
        \vertex (v1) at (0, 0);
        \vertex (o1) at (2, 1.5) {\(\boldsymbol{L_{\mu}}\)};
        \vertex (o2) at (2, -1.5) {\(\boldsymbol{\eta}\)};
        
        \diagram* {
          (i1) -- [plain,very thick] (v1),
          (v1) -- [fermion, very thick] (o1),
          (v1) -- [scalar,very thick] (o2),
        };
      \end{feynman}
    \end{tikzpicture}
    \hspace{1cm} 
       \begin{tikzpicture}[scale=0.6]
      \begin{feynman}
        \vertex (i1) at (-3, 0) {\(\boldsymbol{\psi}\)};
        \vertex (v1) at (0, 0);
        \vertex (loop1) at (2, 0); 
        \vertex (loop2) at (2, 4); 
        \vertex (v2) at (2, 0);
        \vertex (v3) at (4,0);
        \vertex (o1) at (7, 1.2) {\(\boldsymbol{L_{\mu}}\)};
        \vertex (o2) at (7, -1.2) {\(\boldsymbol{\eta}\)};
        
        \diagram* {
          (i1) -- [plain,very thick] (v1) -- [white,plain] (v2)  -- [plain,very thick,edge label=\(\boldsymbol{{N_2}}\)] (v3)-- [fermion,very thick] (o1),
          (v3) -- [scalar,very thick] (o2),
          (v1) -- [scalar,very thick, half left, edge label=\(\boldsymbol{{\eta}}\)] (loop1) -- [fermion, very thick,half left, edge label=\(\boldsymbol{L_{\mu}}\)] (v1) 
        };
      \end{feynman}
    \end{tikzpicture}
    \end{center}
    \caption{\small{Tree level and one loop self-energy diagrams relevant for resonant leptogenesis.}}
    \label{fig:lepto_loop}
\end{figure}


 \begin{align}
 \nonumber
 \label{eq: CPasymmetry}
\epsilon_{N_2} = \frac{\text{Im}\big[(y^{\dagger}_{\eta 2}y_{\psi})^{2}\big]}{(y^{\dagger}_{\eta 2}y_{\eta 2})(y^{\dagger}_{\psi}y_{\psi})}\frac{(M^{2}_2 - m^{2}_{\psi})M_2\Gamma_{\psi}}{(M^{2}_2 - m^{2}_{\psi})^{2}+ M^{2}_2\Gamma^{2}_{\psi}}, 
\\
\epsilon_{\psi} = \frac{\text{Im}\big[(y^{\dagger}_{\psi}y_{\eta 2})^{2}\big]}{(y^{\dagger}_{\eta 2}y_{\eta 2})(y^{\dagger}_{\psi}y_{\psi})}\frac{(m^{2}_{\psi} - M^{2}_{2})m_{\psi}\Gamma_{N_2}}{(m^{2}_{\psi} - M^{2}_{2})^{2}+ m^{2}_{\psi}\Gamma^{2}_{N_2}}.
\end{align}

\noindent The relevant Boltzmann equations for leptogenesis are

\begin{align}\label{Boltz}
\nonumber
\frac{dY_{N_2}}{dz} &= -D_{N_2} (Y_{N_2}-Y^{eq}_{N_2}), \\ \nonumber
\frac{dY_{\psi}}{dz} &= -D_{\psi} (Y_{\psi}-Y^{eq}_{\psi})-S_{A}(Y_{\psi}^{2}-(Y_{\psi}^{eq})^{2}), \\
  \frac{dY_{B-L_{\mu}}}{dz} &= -\epsilon_{N_2} D_{N_2} (Y_{N_2}-Y^{eq}_{N_2})
    -\epsilon_{\psi} D_{\psi} (Y_{\psi}-Y^{eq}_{\psi}) - W_{ID}^{\psi} Y_{B-L_{\mu}}  \\ \nonumber
   &
   - W_{ID}^{N_2} Y_{B-L_{\mu}} - W_{\Delta L} Y_{B-L_{\mu}},
\end{align}
where the decay terms $D_{N_2}$ and $D_{\psi}$ are given by 
\begin{align}
     D_{N_2} = K_{N_2} \left(  z \right) \frac{\kappa_{1}\left(z\right)}{\kappa_{2}\left(z\right)}, \quad
  D_{\psi} = K_{\psi} \left( m_{\psi}/M_{2} z \right) \frac{\kappa_{1}\left(\frac{m_{\psi}}{M_2}z\right)}{\kappa_{2}\left(\frac{m_{\psi}}{M_2}z\right)}   
\end{align}
 and $K_{N_2}$, $K_{\psi}$ is written as 
 \begin{align}
  K_{N_2} =  \frac{\Gamma_{N_2}}{H(T=M_{2})}, \quad
    K_{\psi} =  \frac{\Gamma_{\psi}}{H(T=m_{\psi})}.
 \end{align}

\noindent The decay widths $\Gamma_{N_2}$ and $\Gamma_{\psi}$ are given by
 
 \begin{align}
     \Gamma_{N_2} = \frac{M_{2}}{8 \pi}(y_{\eta_2}^{\dagger}y_{\eta_2})\left(1- \frac{m_{\eta}^{2}}{M_{2}^{2}} \right), \quad
      \Gamma_{\psi} = \frac{m_\psi}{8 \pi}(y_{\psi}^{\dagger}y_{\psi})\left(1- \frac{m_{\eta}^{2}}{m_{\psi}^{2}} \right).
 \end{align}
 The terms $W_{ID}^{N_2}$ and $W_{ID}^{\psi}$ are the washout terms due to the inverse decay of $N_2$ and $\psi$ and are given by

 \begin{align}
  W_{ID}^{N_2} = \frac{1}{4}K_{N_2}(z)^{3}\kappa_{1}\left(z\right), \quad
     W_{ID}^{\psi} = \frac{1}{4}K_{\psi}(m_{\psi}/M_2 z)^{3}\kappa_{1}\left(\frac{m_{\psi}}{M_2}z\right).
 \end{align}
 The term $W_{\Delta L}$ and takes care of the washouts coming from the lepton number violating scattering terms defined as $W_{\Delta L}=\Gamma/H z^{2}$. The important scatterings in this model are $l_{\mu} \eta\longrightarrow \overline{l_{\mu}} \eta^{*}$ and $l_{\mu}l_{\mu}\longrightarrow \eta \eta^{*}$. 
The gauge boson mediated scattering term $S_{A}$ for fermion triplet $\psi$ is given by
\begin{equation}
    S_{A} = \bigg(\frac{\pi^2 g^{*1/2}M_Pl}{1.66*180 g_{\psi}^2}\bigg)\frac{1}{m_{\psi}}\bigg(\frac{I_{z}}{m_{\psi}/M_2z \kappa_{2}(m_{\psi}/M_2z)^{2}}\bigg),
\end{equation}
where the form of $I_{z}$ is given as
\begin{equation}
    I(z) =  \int_{4}^{\infty} \sqrt{x} \kappa (m_{\psi}/M_2z \sqrt{x})\hat{\sigma}_{A}(x)dx .
\end{equation}
Here, $\hat{\sigma}_{A}(x)$ is cross section for gauge boson mediated processes given by \cite{Biswas:2023azl} 
\begin{equation}
\hat{\sigma}_{A}(x) = \frac{6g^2}{72 \pi}\Bigg[\frac{45}{2}r(x)-\frac{27}{2}r(x)^3-\{9(r(x)^2-2)+18(r(x)^2-1)^2\}\text{ln}\bigg(\frac{1+r(x)}{1-r(x)}\bigg)\Bigg],   
\end{equation}
with $r(x) = \sqrt{1-4/x}$.

\subsubsection{Leptogenesis results and discussion}
\label{sec:results}

\begin{figure}[htb!]
     \centering
     \includegraphics[scale=0.5]{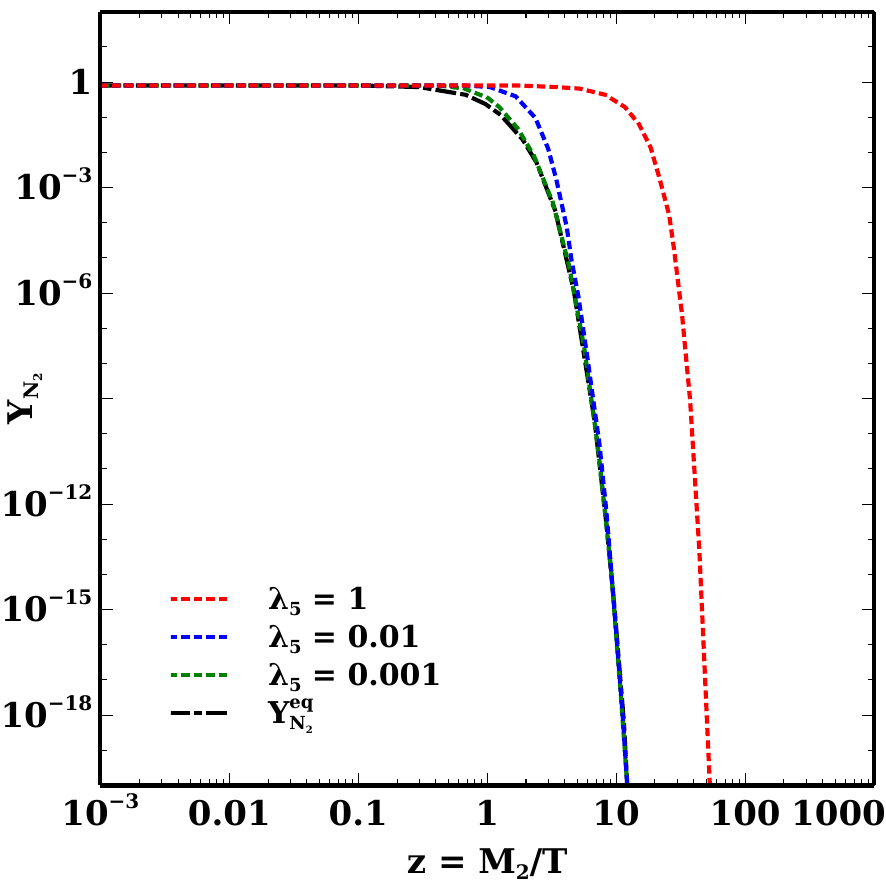}
\includegraphics[scale=0.5]{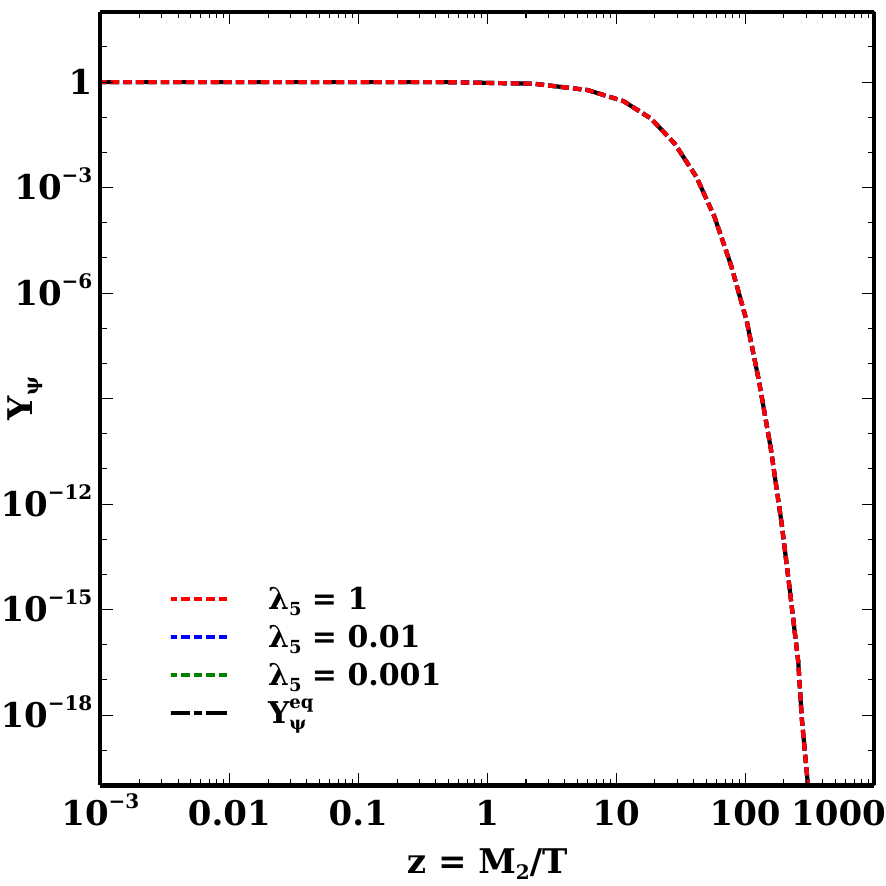}
\includegraphics[scale=0.5]{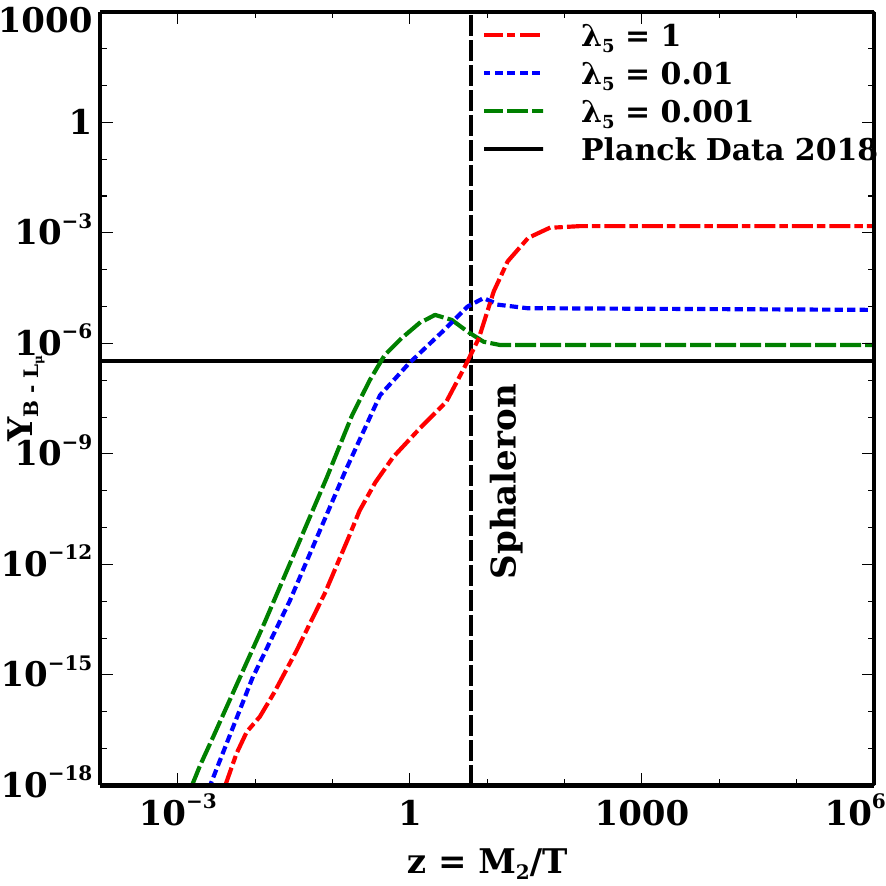}
     \caption{\small{Variation of co moving number density $Y_{N_2}$ (upper left panel), $Y_{\psi}$ (upper right panel) and $B-L_{\mu}$ asymmetry (lower panel) with $z=M_{2}/T$ for different benchmark values of the quartic coupling $\lambda_{5}$.
   The other parameters are fixed at $M_1 = 7.9\times  10^{-1}$ TeV,  $M_2 =8 \times 10^{-1}$ TeV, $M_3 = 9 \times 10^{-1}$ TeV, $m_{\eta} = 7.9\times 10^{-1}$ TeV, $m_{1} =10^{-13} $ eV, $m_{\psi} = 8\times 10^{-1}$ TeV and $y_{\psi} = 2\times 10^{-3}$. The horizontal lines in right panel plot depicts the required $B-L_{\mu}$ asymmetry to generate net baryon asymmetry of the Universe (Planck Data 2018) after sphaleron transition (Vertical dotted line).}}
    \label{fig2}
 \end{figure}

This section discusses the results of resonant leptogenesis from fermion triplet $\psi$ and $N_2$. 
By constraining the mass of triplet $m_{\psi}$ and Yukawa coupling $y_{\psi}$ from dark matter and Muon ($g-2$) analysis, we solved the Boltzmann equations Eq. (\ref{Boltz}) for leptogenesis. In Fig. \ref{fig2}, we show the evolution of co-moving number density of $N_2$, $\psi$ and ${B-L_{\mu}}$ with $z = M_{2}/T$ for different benchmark values of $\lambda_5$. The upper left panel plot shows that with an increase in $\lambda_5$, the $N_2$ deviates more from its equilibrium. The quartic coupling $\lambda_{5}$ and the Yukawa couplings $y_{\eta_2}$ contributes to the neutrino mass generation. With the increase in $\lambda_{5}$ the required
 value of the Yukawa couplings $Y_{\eta_2} $ decreases. This lead to a decrease in the decay and inverse decay process resulting in a larger deviation from the equilibrium abundance.  From the upper right panel plot, it can be seen that changes in the quartic coupling $\lambda_5$ does not affect the co-moving number density of $\psi$ and it tracks its equilibrium abundance. It is because, $\psi$ abundance is mainly determined by decay term $D_{\psi}$ and scattering term $S_{A}$, both of which are independent of $\lambda_5$. On the lower panel plot it can be seen that with the increase in $\lambda_{5}$ the $B-L_{\mu}$ asymmetry increases. Since, an increase in $\lambda_{5}$, leads to a larger deviation from equilibrium abundance for $N_{2}$, it results in an increase in the production of asymmetry. 

  \begin{figure}[htb!]
     \centering
     \includegraphics[scale=0.5]{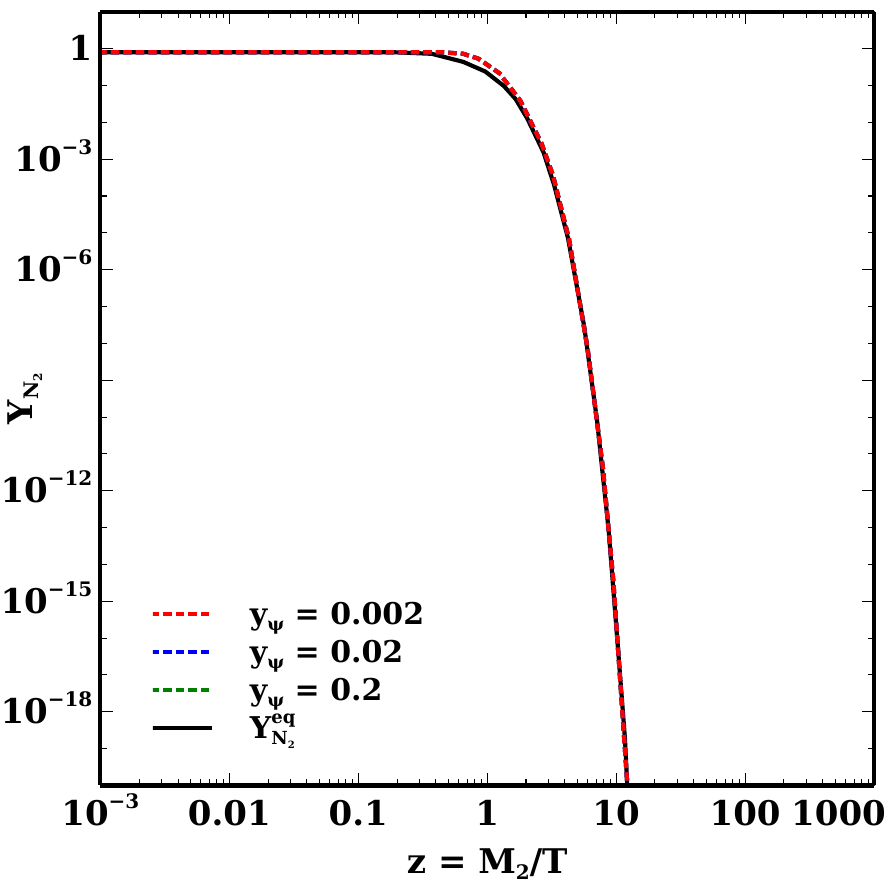}
\includegraphics[scale=0.5]{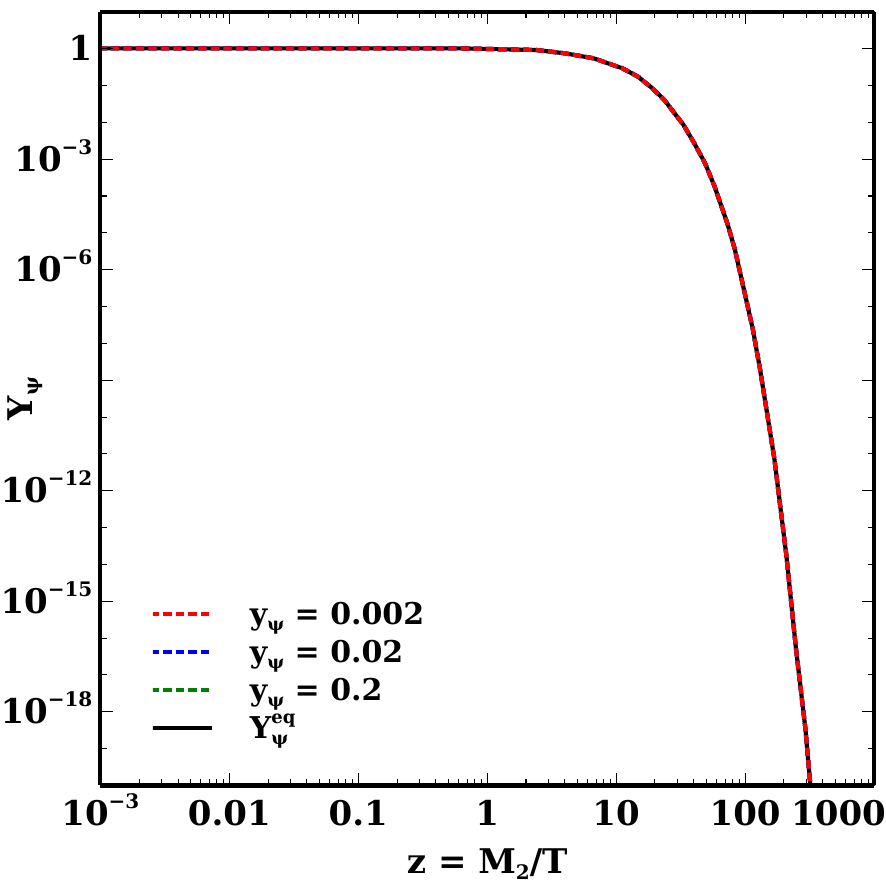}
\includegraphics[scale=0.5]{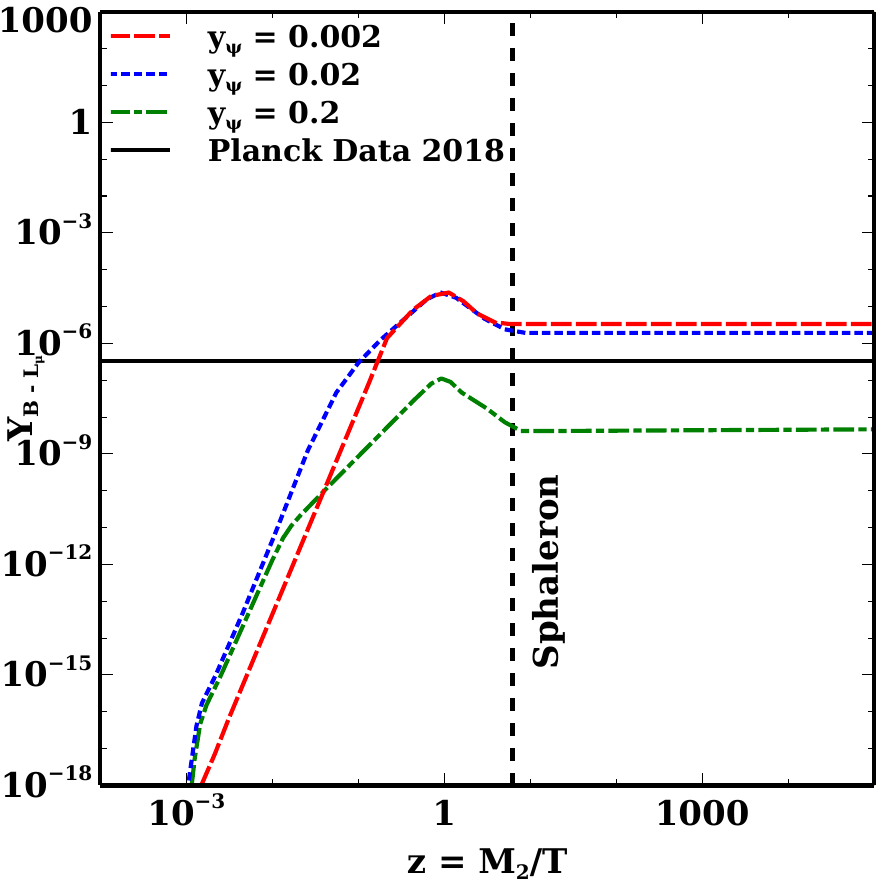}
    \caption{\small{Variation of co moving number density $Y_{N_2}$ (upper left panel), $Y_{\psi}$ (upper right panel) and $B-L_{\mu}$ asymmetry (lower panel) with $z=M_{2}/T$ for different benchmark values of the Yukawa couplings $y_{\psi}$.
    The other parameters are fixed at $M_1 = 7.9 \times 10^{-1}$ TeV,  $M_2 = 0.8$ TeV, $\Delta m_{N_2\psi} = 10^{-10}$ TeV, $M_3 = 0.9$ TeV, $m_{\eta} = 7.9\times 10^{-1}$ TeV, $m_{1} =10^{-13}$ eV, $m_{\psi} = 0.8$ TeV and $\lambda_{5} = 10^{-3}$. The horizontal lines in right panel plot depicts the required $B-L_{\mu}$ asymmetry to generate net baryon asymmetry of the Universe (Planck Data 2018) after the sphaleron transition (Vertical dotted line).}}
     \label{fig3}
 \end{figure}

 In Fig. \ref{fig3} we show the evolution of $N_{2}$, $\psi$ and $B-L_{\mu}$ asymmetry with $z=M_{2}/T$ with different benchmark values of $y_{\psi}$. In the upper left panel plot, we see that the $N_{2}$ nearly tracks its equilibrium abundance and the abundance does not change with the change in $y_{\psi}$. With the chosen $\lambda_{5}=10^{-3}$, the Yukawa couplings $y_{\eta 2}$ are large enough to keep the $N_{2}$ close to its equilibrium abundance. As the $y_{\psi}$ is not involved in the decay or inverse decay of $N_{2}$, its abundance remains constant with the change in $y_{\psi}$. The top right panel shows that the $\psi$ tracks its equilibrium irrespective of $y_{\psi}$. Although, a change in $y_{\psi}$ changes the decay and inverse decay rate of $\psi$, its gauge annihilation always keeps it near equilibrium. In the lower panel plot of Fig. \ref{fig3} it is observed that with the increase in $y_{\psi}$ the $B-L_{\mu}$ asymmetry decreases. As $y_{\psi}$ increases, the decay width of $\psi$ as well as the CP asymmetry parameters $\epsilon_{N_{2}}$ and $\epsilon_{\psi}$ increase. However, it also results in strong washouts from the inverse decays and scatterings, decreasing $B-L_{\mu}$ asymmetry. The net lepton asymmetry generated in the muon sector is converted into baryon asymmetry via sphaleron factor $Y_{B} = c_{sph}Y_{B-L_{\mu}}$, where $c_{sph} = 33/57$ for our model. The details are given in Appendix (\ref{appen2}).

 In Fig. \ref{fig:scan}, we show a scan plot in $m_{\psi}$ vs $y_{\psi}$ plane (left) and $m_{\eta}$ vs $\Delta m$ plane (right) that satisfies Muon ($g-2$), observed baryon asymmetry and DM relic. In the left panel plot, the cyan dot markers satisfy the observed Muon $(g-2)$, while the red star markers satisfy the observed baryon asymmetry as well as Muon $(g-2)$. The blue star markers satisfy the observed DM relic, direct detection bounds, observed BAU and observed excess of Muon $(g-2)$. In the right panel plot, the cyan dot markers satisfy the observed Muon $(g-2)$, while the red dot markers satisfy observed DM relic and direct detection bounds. The blue dot markers satisfy observed BAU also. The blue star points are also given in table (\ref{tab2}).

\begin{figure}[htb!]
	\centering
	\includegraphics[scale=0.5]{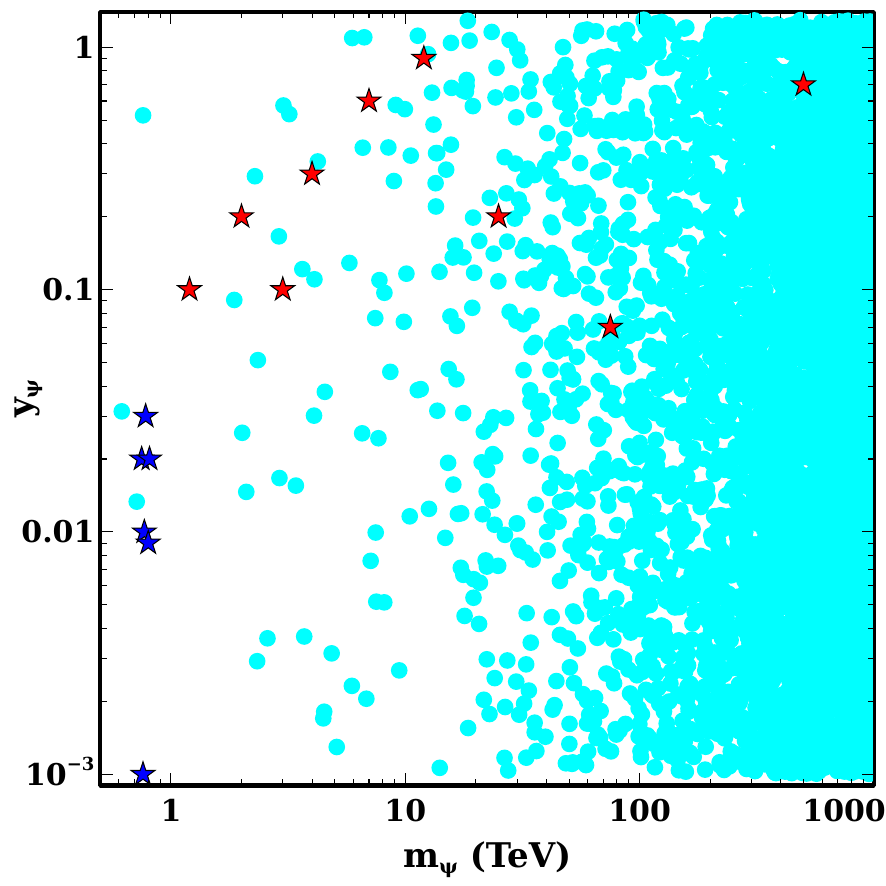}
    \includegraphics[scale=0.5]{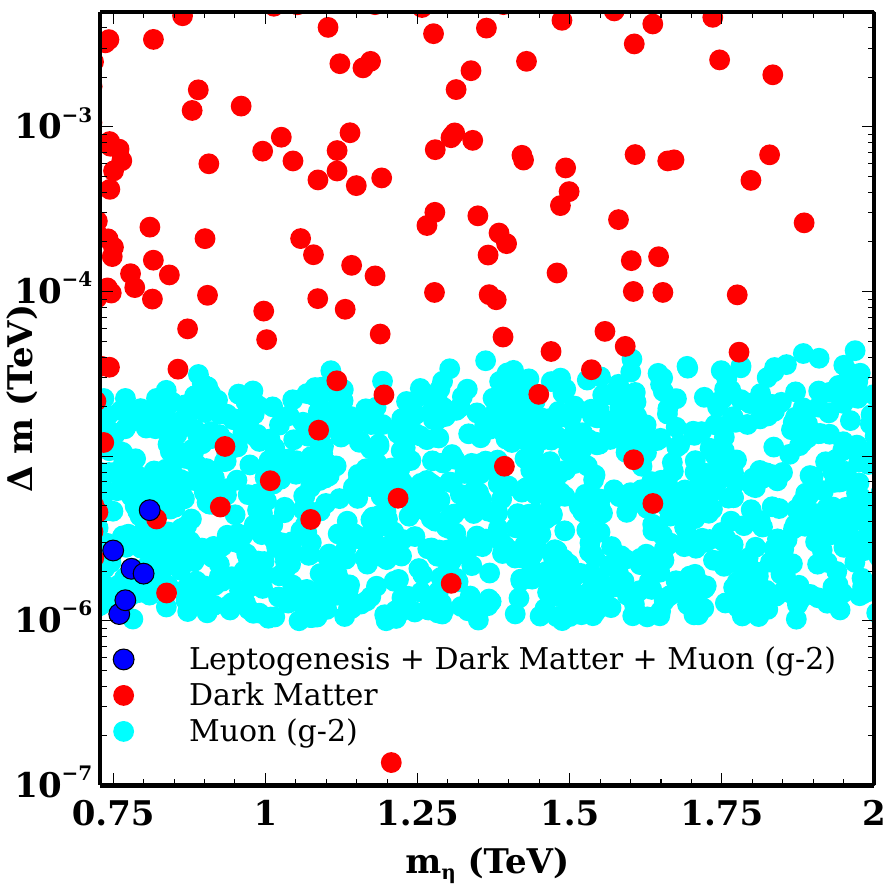}
	\caption{\small{Scan plot in $m_{\psi} - y_{\psi}$ plane (left). Here cyan points satisfy Muon ($g-2$) and the red stars represent points that satisfy Muon ($g-2$) and observed BAU. The blue star points satisfy DM relic, direct detection limit, BAU and FNL results on Muon ($g-2$). Parameter space in $m_{\eta} - \Delta m$ plane (right). Here cyan points satisfy Muon ($g-2$), red points satisfy dark matter and direct detection limit and blue data points satisfy BAU along with dark matter constraints and Muon $(g-2)$ observed excess.  }}
	\label{fig:scan}
\end{figure}

 \begin{table}[htb!]
     \centering
    
     \begin{tabular}{|c|c|c|c|c|c|c| }
      \hline \hline
   $m_{\psi}$ (TeV) & $m_{\eta}$ (TeV)   &$y_{\psi}$ & $M_{2}$ (TeV)&$\lambda_{5}$ & $\epsilon_{\psi}$& $\epsilon_{N_2}$\\ \hline
   0.75 & 0.749997  &  0.02 &  $0.7500000005$ & $0.009$ &$3.5 \times 10^{-7}$& 0.094
   \\ \hline
   0.76 & 0.759999  &  0.001 &  0.7600001 & 0.03 &$1.9 \times10^{-10}$&$3.9\times 10^{-7}$
   \\ \hline
   0.77 & 0.769999  & 0.01 &  0.7700001 & 0.02 &$2.9\times10^{-10}$&$3.9\times 10^{-5}$
   \\ \hline
   0.78 & 0.779998  &  0.03 &  0.7800000009 & 0.003 &$4\times10^{-7}$&0.078
   \\ \hline
   
   0.8 & 0.799998  & 0.009 &  $0.800000001$ & $0.002$ &$5.63 \times 10^{-7}$&0.0064
   \\ \hline
   0.81 & 0.809995  & 0.02 &  0.8100000006 & $0.03$ &$1.58 \times 10^{-7}$&0.129
\\ \hline\hline
     \end{tabular}
     \caption{\small{Benchmark points that satisfy leptogenesis, dark matter and muon ($g-2$) anomaly simultaneously.}}
     \label{tab2}
 \end{table}
 
\section{CONCLUSIONS}

\label{sec:conclusion}

We propose a minimal extension of the scotogenic model that has the potential to explain nonzero neutrino mass, muon ($g-2$) along with dark matter and baryon asymmetry of the Universe. Three copies of Majorana right-handed neutrinos generate the light neutrino masses through the scotogenic mechanism. Because of the imposed $Z_{4}$ symmetry, the $Z_{2}$ odd fermion triplet can provide a one-loop contribution to the muon's anomalous magnetic moment. The fermion triplet can also generate net lepton asymmetry in the muon sector through its out-of-equilibrium decay. The Yukawa coupling involved in leptogenesis is free from neutrino sector. The strong gauge annihilation keeps the triplet fermion abundance close to its equilibrium abundance. The Yukawa couplings of $N_{2}$ are large enough and undergo strong washouts by the inverse decays. Both these factors reduce the lepton asymmetry production. It leads to a high-scale leptogenesis around $\sim 10^{10}$ GeV. To successfully generate the lepton asymmetry around the TeV scale, we consider the resonance enhancement of asymmetry between $\psi$ and $N_{2}$. The Yukawa coupling involved in leptogenesis $y_{\psi}$, also contributes to the muon $(g-2)$ and provides a strong correlation between the two. The coannihilation of the DM, $\eta$ with the triplet fermion $\psi$ is significant in getting the correct DM relic. The mass difference between the triplet fermion and the doublet scalar becomes a crucial parameter. Both muon $(g-2)$ and DM relic favors toward a parameter space where $ m_{\psi}\sim m_{DM}$. The requirement of correct asymmetry at the TeV scale necessarily imposes the resonance condition $M_{2} - m_{\psi} \sim \Gamma_{2,\psi}/2$.

 As the scale of triplet fermion involved in leptogenesis is in the TeV range, the model is testable in future LHC experiments. The ATLAS experiment at LHC has obtained a lower limit on the mass of heavy fermion triplets around 750 GeV at 95\% confidence level \cite{ATLAS:2020wop}. The phenomenology of inert scalar doublet as dark matter is also discussed in the paper. We constrained our model by the recent relic density results and direct detection bounds. We show that rich phenomenology among the muon $(g-2)$ observed excess, leptogenesis and DM production around the TeV scale. Such low-scale models may be tested near future experiments.

\vspace{0.5 cm}
\noindent
\vspace{0.5 cm}
\textbf{\Large{Acknowledgments}}

\noindent B.C.C. is thankful to the Inter-University Centre for Astronomy and Astrophysics (IUCAA) for providing the necessary facilities during the completion of this work.

\appendix
\section{CROSS SECTIONS FOR THE DARK MATTER COANNIHILATION  }\label{appen1}
Here we show the cross sections of the dark matter coannihilation,
\begin{align}
\sigma _{\eta \psi \longrightarrow L_{\mu} h} &= -\frac{1}{16 \pi \left(-4 m_{\psi}^2 m_{\eta}^2 + \left(-m_{\psi}^2 - m_{\eta}^2 + s\right)^2\right)} \bigg[ \notag \\
&\quad \left( -\left((\lambda_3 + \lambda_4 + \lambda_5)v\right)^2 \left( \frac{-y_{\psi}}{2} \right) \left( \frac{-y_{\psi}}{2} \right) \right)  \notag \\
&\quad \times \frac{- \left(4 (m_{\psi}^2 - m_{\eta}^2 + 2 m_{\psi} m_{\mu} + m_{\mu}^2) s^3 \right)}{\sqrt{\frac{1}{s^4} \left(m_{\psi}^4 + (m_{\eta}^2 - s)^2 - 2 m_{\psi}^2 (m_{\eta}^2 + s)\right) \left(m_{h}^4 + (m_{\mu}^2 - s)^2 - 2 m_{h}^2 (m_{\mu}^2 + s)\right)}} \notag \\
&\quad \times \frac{1}{(m_{\eta}^2 m_{\mu}^2 - m_{\eta}^2 s + m_{\mu}^2 s - s^2 + m_{h}^2 (-m_{\eta}^2 + s) + m_{\psi}^2 (m_{h}^2 - m_{\mu}^2 + s))} \notag \\
&\quad \times \frac{1}{(s^2 \sqrt{\frac{1}{s^4} \left(m_{\psi}^4 + (m_{\eta}^2 - s)^2 - 2 m_{\psi}^2 (m_{\eta}^2 + s)\right) \left(m_{h}^4 + (m_{\mu}^2 - s)^2 - 2 m_{h}^2 (m_{\mu}^2 + s)\right)})} \notag \\
&\quad + \log\left(\frac{m_{\eta}^2 m_{\mu}^2 - m_{\eta}^2 s + m_{\mu}^2 s - s^2 + m_{h}^2 (-m_{\eta}^2 + s) + m_{\psi}^2 (m_{h}^2 - m_{\mu}^2 + s)}{s^2 \sqrt{\frac{1}{s^4} \left(m_{\psi}^4 + (m_{\eta}^2 - s)^2 - 2 m_{\psi}^2 (m_{\eta}^2 + s)\right) \left(m_{h}^4 + (m_{\mu}^2 - s)^2 - 2 m_{h}^2 (m_{\mu}^2 + s)\right)}}\right) \notag \\\notag
&\quad - \log\left(\frac{m_{\eta}^2 m_{\mu}^2 - m_{\eta}^2 s + m_{\mu}^2 s - s^2 + m_{h}^2 (-m_{\eta}^2 + s) + m_{\psi}^2 (m_{h}^2 - m_{\mu}^2 + s) - s^2}{\sqrt{\frac{1}{s^4} \left(m_{\psi}^4 + (m_{\eta}^2 - s)^2 - 2 m_{\psi}^2 (m_{\eta}^2 + s)\right) \left(m_{h}^4 + (m_{\mu}^2 - s)^2 - 2 m_{h}^2 (m_{\mu}^2 + s)\right)}}\right) \bigg], 
\\\notag
         \sigma_{\psi \eta \longrightarrow N_{2}\eta} &= \frac{1}{16 \pi (-4 m_{\psi}^2 m_{\eta}^2 + (-m_{\psi}^2 - m_{\eta}^2 + s)^2)}\Bigg[ \frac{1}{(m_{\mu}^{2}-s)^2}
    \\\quad\nonumber\times
    &\sqrt{\frac{1}{s^{4}}((m_{\psi}^4 + (m_{\eta}^2 - s)^2 - 
   2 m_{\psi}^2 (m_{\eta}^2 + s)) (m_{\eta}^4 + (M_{2}^2 - s)^2 - 
   2 m_{\eta}^2 (M_{2}^2 + s)))} \\\quad\nonumber\times& \bigg(4 m_{\psi} s (-m_{\eta}^2 m_{\mu} + (M_{2} + m_{\mu}) (M_{2} m_{\mu} + s)) + (m_{\eta}^2- 
    s) (-4 M_{2} m_{\mu} s + m_{\eta}^2 (m_{\mu}^2 + s) \\\quad\nonumber &- M_{2}^2 (m_{\mu}^2 + s) - 
    s (m_{\mu}^2 + s)) + 
 m_{\psi}^2 (4 M_{2} m_{\mu} s - m_{\eta}^2 (m_{\mu}^2 + s) + M_{2}^2 (m_{\mu}^2 + s) \\\quad\nonumber&+ 
    s (m_{\mu}^2 + s))\bigg)
    \times \bigg(\frac{-y_{\psi}}{\sqrt{2}}\bigg)^{2}\bigg(\frac{-y_{\eta_2}}{\sqrt{2}}\bigg)^{2} + 
    \frac{1}{(m_{\mu}^{2}-s)^{2}}  \\\quad \nonumber & \times
\sqrt{\frac{1}{s^4}((m_{\psi}^4 + (m_{\eta}^2 - s)^2 - 
   2 m_{\psi}^2 (m_{\eta}^2 + s)) (m_{\eta}^4 + (M_{2}^2 - s)^2 - 
   2 m_{\eta}^2 (M_{2}^2 + s)))} \\\quad\nonumber&\times 
   \bigg(4 m_{\psi} s (-m_{\eta}^2 m_{\mu} + (M_{2} + m_{\mu}) (M_{2} m_{\mu} + s)) + (m_{\eta}^2 - 
     s) (-4 M_{2} m_{\mu} s + m_{\eta}^2 (m_{\mu}^2 + s)\\\quad\nonumber&- M_{2}^2 (m_{\mu}^2 + s) - 
     s (m_{\mu}^2 + s)) + 
  m_{\psi}^2 (4 M_{2} m_{\mu} s - m_{\eta}^2 (m_{\mu}^2 + s) + M_{2}^2 (m_{\mu}^2 + s) + 
     s (m_{\mu}^2 + s))\bigg)  \\\quad\nonumber &\times
  \bigg(  \frac{-y_{\psi}}{\sqrt{2}} \frac{-y_{\eta_2}}{\sqrt{2}} \frac{-y_{\psi}}{\sqrt{2}} \frac{-y_{\eta_2}}{\sqrt{2}}\bigg) + 2\bigg( \frac{-y_{\psi}}{\sqrt{2}} \frac{-y_{\eta_2}}{\sqrt{2}} \frac{-y_{\psi}}{\sqrt{2}} \frac{-y_{\eta_2}}{\sqrt{2}} \bigg) \\\nonumber\quad& \times
    \Bigg(-\bigg(4 (m_{\psi}^2 - m_{\eta}^2 + 2 m_{\psi} m_{\mu} + 
   m_{\mu}^2) (m_{\eta}^2 - (M_{2} + m_{\mu})^2) s^3\\\nonumber\quad& \times \sqrt{\frac{1}{s^4}(m_{\psi}^4 + (m_{\eta}^2 - s)^2 - 
      2 m_{\psi}^2 (m_{\eta}^2 + s)) (m_{\eta}^4 + (M_{2}^2 - s)^2 - 
      2 m_{\eta}^2 (M_{2}^2 + s))}/
      \\\quad\nonumber& \bigg(-m_{\eta}^4 + m_{\eta}^2 (M_{2}^2 - 2 s) + m_{\psi}^2 (m_{\eta}^2 - M_{2}^2 - s) + 
   s (-M_{2}^2 + 2 m_{\mu}^2 + s \\\quad\nonumber&- 
      s \sqrt{\frac{1}{s^4}
         (m_{\psi}^4 + (m_{\eta}^2 - s)^2 - 
            2 m_{\psi}^2 (m_{\eta}^2 + s)) (m_{\eta}^4 + (M_{2}^2 - s)^2 - 
            2 m_{\eta}^2 (M_{2}^2 + s)))}\bigg) \\
            \end{align}
            
            \begin{align}\nonumber
            \\\quad\nonumber&(-m_{\eta}^4 + m_{\eta}^2 (M_{2}^2 - 2 s) +
    m_{\psi}^2 (m_{\eta}^2 - M_{2}^2 - s) + 
   s (-M_{2}^2 + 2 m_{\mu}^2 + s \\\quad\nonumber&+ 
      s \sqrt{\frac{1}  {s^4}(m_{\psi}^4 + (m_{\eta}^2 - s)^2 - 
            2 m_{\psi}^2 (m_{\eta}^2 + s)) (m_{\eta}^4 + (M_{2}^2 - s)^2 - 
            2 m_{\eta}^2 (M_{2}^2 + s))}))\bigg)
    \\\quad\nonumber& +(m_{\psi}^2 + M_{2}^2 + 2 M_{2} m_{\mu} + m_{\mu}^2 + 2 m_{\psi} (M_{2} + m_{\mu}) - s)
    \\\quad\nonumber& \times\log\bigg[
m_\eta^4 - m_\eta^2 (M_2^2 - 2s) + m_\psi^2 (-m_\eta^2 + M_2^2 + s)
+ s \bigg( M_2^2 - 2m_\mu^2 + s( -1 \\\quad\nonumber+& 
\sqrt{\frac{1}{s^4} \big( m_\psi^4 + (m_\eta^2 - s)^2 - 2m_\psi^2 (m_\eta^2 + s) \big)
\big( m_\eta^4 + (M_2^2 - s)^2 - 2m_\eta^2 (M_2^2 + s) \big)}) \bigg)/
 \\\quad\nonumber & m_\eta^4- m_\eta^2 (M_2^2 - 2s) + m_\psi^2 (-m_\eta^2 + M_2^2 + s) 
- s \big( -M_2^2 + 2m_\mu^2 + s  
\\\quad\nonumber&+ s \sqrt{\frac{1}{s^4} ( m_\psi^4 + (m_\eta^2 - s)^2 - 2m_\psi^2 (m_\eta^2 + s)) 
( m_\eta^4 + (M_2^2 - s)^2 - 2m_\eta^2 (M_2^2 + s))} \big)
\bigg] 
     \Bigg)
    \\\quad\nonumber -&(1/(m_{\mu}^2 - 
   s)) 4 \bigg(\frac{-y_{\psi}}{\sqrt{2}} \frac{-y_{\eta_2}}{\sqrt{2}} \frac{-y_{\psi}}{\sqrt{2}} \frac{-y_{\eta_2}}{\sqrt{2}}\bigg) (s (m_{\psi} m_{\mu} + M_{2} m_{\mu} + m_{\mu}^2+s) \\\quad\nonumber&  \times \sqrt{\frac{1}{s^4}(m_{\psi}^4 + (m_{\eta}^2 - s)^2 - 
   2 m_{\psi}^2 (m_{\eta}^2 + s)) (m_{\eta}^4 + (M_{2}^2 - s)^2 - 
   2 m_{\eta}^2 (M_{2}^2 + s))} 
   \\\quad\nonumber+& \bigg(-m_{\eta}^4 + m_{\psi}^3 M_{2} - 2 m_{\eta}^2 (m_{\psi} + m_{\mu}) (M_{2} + m_{\mu}) + 
  m_{\psi}^2 M_{2} (M_{2} + 2 m_{\mu})\nonumber\\\quad\nonumber+& 
  m_{\psi} (M_{2}^3 + 2 M_{2}^2 m_{\mu} + 2 M_{2} m_{\mu}^2 + m_{\mu}^3 + m_{\mu} s) + 
  m_{\mu} (m_{\mu}^3 + 2 m_{\mu} s + M_{2} (m_{\mu}^2 + s))\bigg)
  \\\quad\nonumber& \log\bigg[(m_{\eta}^4 - m_{\eta}^2 (M_{2}^2 - 2 s) + m_{\psi}^2 (-m_{\eta}^2 + M_{2}^2 + s) + 
      s (M_{2}^2 - 2 m_{\mu}^2 + 
         s (-1 
         \\\quad\nonumber+& \sqrt{\frac{1}{s^4}(m_{\psi}^4 + (m_{\eta}^2 - s)^2 - 
                 2 m_{\psi}^2 (m_{\eta}^2 + s)) (m_{\eta}^4 + (M_{2}^2 - s)^2 - 
                 2 m_{\eta}^2 (M_{2}^2 + s))})))/\\\quad\nonumber&(m_{\eta}^4 - 
      m_{\eta}^2 (M_{2}^2 - 2 s) + m_{\psi}^2 (-m_{\eta}^2 + M_{2}^2 + s) - 
      s (-M_{2}^2 + 2 m_{\mu}^2 + s 
      \\\quad\nonumber +& 
         s \sqrt{\frac{1}
            {s^4}(m_{\psi}^4 + (m_{\eta}^2 - s)^2 - 
               2 m_{\psi}^2 (m_{\eta}^2 + s)) (m_{\eta}^4 + (M_{2}^2 - s)^2 - 
               2 m_{\eta}^2 (M_{2}^2 + s))}))\bigg]) 
               \\\quad\nonumber -& s (m_{\psi} m_{\mu} + M_{2} m_{\mu} + m_{\mu}^2 + s) \\\quad\nonumber&\times \sqrt{\frac{1}{s^4}(m_{\psi}^4 + (m_{\eta}^2 - s)^2 - 
       2 m_{\psi}^2 (m_{\eta}^2 + s)) (m_{\eta}^4 + (M_{2}^2 - s)^2 - 
       2 m_{\eta}^2 (M_{2}^2 + s))} 
       \\\quad\nonumber+& \big(-m_{\eta}^4 + m_{\psi}^3 M_{2} - 
    2 m_{\eta}^2 (m_{\psi} + m_{\mu}) (M_{2} + m_{\mu}) + m_{\psi}^2 M_{2} (M_{2} + 2 m_{\mu}) + 
    m_{\psi} (M_{2}^3 \\\quad\nonumber+& 2 M_{2}^2 m_{\mu} + 2 M_{2} m_{\mu}^2 + m_{\mu}^3 + m_{\mu} s) + 
    m_{\mu} (m_{\mu}^3 + 2 m_{\mu} s + M_{2} (m_{\mu}^2 + s))\big)\\\quad\nonumber& \log\bigg[(m_{\eta}^4 - 
      m_{\eta}^2 (M_{2}^2 - 2 s) + m_{\psi}^2 (-m_{\eta}^2 + M_{2}^2 + s) + 
      s (M_{2}^2 - 2 m_{\mu}^2 \\\quad\nonumber+& 
         s (-1 + \sqrt{\frac{1}{
              s^4}(m_{\psi}^4 + (m_{\eta}^2 - s)^2 - 
                 2 m_{\psi}^2 (m_{\eta}^2 + s)) (m_{\eta}^4 + (M_{2}^2 - s)^2 - 
                 2 m_{\eta}^2 (M_{2}^2 + s))})))/\\\quad\nonumber&(m_{\eta}^4 - 
      m_{\eta}^2 (M_{2}^2 - 2 s) + m_{\psi}^2 (-m_{\eta}^2 + M_{2}^2 + s) - 
      s (-M_{2}^2 + 2 m_{\mu}^2 + s 
      \\\quad\nonumber+& 
         s \sqrt{\frac{1}{s^4}
        (m_{\psi}^4 + (m_{\eta}^2 - s)^2 - 
               2 m_{\psi}^2 (m_{\eta}^2 + s)) (m_{\eta}^4 + (M_{2}^2 - s)^2 - 
               2 m_{\eta}^2 (M_{2}^2 + s))}))\bigg]
    \Bigg]
\end{align}
\begin{align}\nonumber
    \sigma_{\eta \psi \rightarrow L_{\mu}Z} &= \frac{1}{16 \pi (-4 m_{\psi}^2 m_{\eta_R}^2 + (-m_{\psi}^2 - 
       m_{\eta_R}^2 + s)^2) }\bigg[EE^2(-1/(m_{\mu}^2 - s)^22 (m_{Z}^2 - m_{\nu}^2 - s)   \\\nonumber \quad &\times \sqrt{\frac{1}{s^4}(m_{\psi}^4 + (m_{\eta_R}^2 - s)^2 - 
   2 m_{\psi}^2 (m_{\eta_R}^2 + s)) (m_{Z}^4 + (m_{\nu}^2 - s)^2 - 
   2 m_{Z}^2 (m_{\nu}^2 + s))}\\\nonumber \quad & \times (4 m_{\psi} m_{\mu} s + m_{\psi}^2 (m_{\mu}^2 + s) - (m_{\eta_R}^2 - s) (m_{\mu}^2 + s)) \\\nonumber \quad & \times  \bigg(-0.25\frac{cos_w}{sin_w} + \frac{sin_w}{cos_w}\bigg)^2 \bigg(\frac{-y_{\psi}}{\sqrt{2}} \frac{-y_{\psi}}{\sqrt{2}}\bigg) - \bigg(\frac{-y_{\psi}}{\sqrt{2}} \frac{y_{\psi}}{\sqrt{2}} \bigg)\bigg(\frac{cos_{w}^{2}+sin_{w}^{2}}{2 cos_{w}sin_{w}}\bigg)^2 \\\nonumber \quad & \times
   ((2 s \sqrt{\frac{1}
      {s^4}(m_{\psi}^4 + (m_{\eta_R}^2 - s)^2 - 
        2 m_{\psi}^2 (m_{\eta_R}^2 + s)) (m_{Z}^4 + (m_{\mu}^2 - s)^2 - 
        2 m_{Z}^2 (m_{\mu}^2 + s))}\\\nonumber \quad & \times
   (4 m_{\psi} (-2 m_{\eta_I}^2 - 2 m_{\eta_R}^2 + m_{Z}^2) m_{\mu} s^2 + 
  m_{\psi}^4 (m_{Z}^2 - m_{\mu}^2 + s)^2 \\\nonumber \quad &+m_{\eta_R}^4 (-m_{Z}^2 + m_{\mu}^2 + s)^2 - 
 2 m_{\eta_R}^2 s (m_{Z}^4 - m_{\mu}^4 - 2 m_{\eta_I}^2 (m_{Z}^2 - m_{\mu}^2) - 2 m_{Z}^2 s + 
    2 m_{\mu}^2 s + s^2) \\\nonumber \quad &+ s^2 (8 m_{\eta_I}^4 + m_{Z}^4 + m_{\mu}^4 + m_{Z}^2 (4 m_{\mu}^2 - 2 s) - 2 m_{\mu}^2 s +
    s^2 + m_{\eta_I}^2 (-6 m_{Z}^2 - 8 m_{\mu}^2 + 4 s) \\\nonumber \quad &- 1/s^2(m_{\psi}^4 + (m_{\eta_R}^2 - s)^2 - 
   2 m_{\psi}^2 (m_{\eta_R}^2 + s)) (m_{Z}^4 + (m_{\mu}^2 - s)^2 \\\nonumber \quad &- 2 m_{Z}^2 (m_{\mu}^2 + s))) - 
 2 m_{\psi}^2 (m_{\eta_R}^2 (m_{Z}^4 - 2 m_{Z}^2 m_{\mu}^2 + m_{\mu}^4 + s^2) + \\\nonumber & +s (-m_{Z}^4 + (m_{\mu}^2 - s)^2 - m_{Z}^2 s + 
      2 m_{\eta_I}^2 (m_{Z}^2 - m_{\mu}^2 + 2 s)))))/
      \\\nonumber \quad & ((m_{\psi}^2 (m_{Z}^2 - m_{\mu}^2 + s) + m_{\eta_R}^2 (-m_{Z}^2 + m_{\mu}^2 + s) + \\\nonumber &
    s (-2 m_{\eta_I}^2 + m_{Z}^2 + m_{\mu}^2 - s \\\nonumber \quad &- 
       s \sqrt{\frac{1}{s^4}
          (m_{\psi}^4 + (m_{\eta_R}^2 - s)^2 - 
             2 m_{\psi}^2 (m_{\eta_R}^2 + s)) (m_{Z}^4 + (m_{\mu}^2 - s)^2 - 
             2 m_{Z}^2 (m_{\mu}^2 + s))}))\\\nonumber \quad & \times (m_{\psi}^2 (m_{Z}^2 - m_{\mu}^2 + s) + 
    m_{\eta_R}^2 (-m_{Z}^2 + m_{\mu}^2 + s) + 
    s (-2 m_{\eta_I}^2 + m_{Z}^2 + m_{\mu}^2 - s \\\nonumber \quad & + 
       s \sqrt{\frac{1}{s^4}(m_{\psi}^4 + (m_{\eta_R}^2 - s)^2 - 
             2 m_{\psi}^2 (m_{\eta_R}^2 + s)) (m_{Z}^4 + (m_{\mu}^2 - s)^2 - 
             2 m_{Z}^2 (m_{\mu}^2 + s))}))) \\\nonumber & +(2 m_{\psi}^2 - 4 m_{\eta_I}^2 - 2 m_{\eta_R}^2 + m_{Z}^2 + 4 m_{\psi} m_{\mu} + 
    2 m_{\mu}^2)\\\nonumber \quad & \times \log\bigg[(m_{\psi}^2 (m_{Z}^2 - m_{\mu}^2 + s) + 
     m_{\eta_R}^2 (-m_{Z}^2 + m_{\mu}^2 + s) + 
     s (-2 m_{\eta_I}^2 + m_{Z}^2 + m_{\mu}^2 - s \\\nonumber \quad &- 
        s \sqrt{\frac{1}{s^4}(m_{\psi}^4 + (m_{\eta_R}^2 - s)^2 - 
              2 m_{\psi}^2 (m_{\eta_R}^2 + s)) (m_{Z}^4 + (m_{\mu}^2 - s)^2 - 
              2 m_{Z}^2 (m_{\mu}^2 + s))}))/\\\nonumber & (m_{\psi}^2 (m_{Z}^2 - m_{\mu}^2 + s) + m_{\eta_R}^2 (-m_{Z}^2 + m_{\mu}^2 + s) + 
  s (-2 m_{\eta_I}^2 + m_{Z}^2 + m_{\mu}^2 - s  \\\nonumber &+ s \sqrt{\frac{1}{s^4}(m_{\psi}^4 + (m_{\eta_R}^2 - s)^2 - 
           2 m_{\psi}^2 (m_{\eta_R}^2 + s)) (m_{Z}^4 + (m_{\mu}^2 - s)^2 - 
           2 m_{Z}^2 (m_{\mu}^2 + s))}))\bigg]) \\\nonumber &+ 1/(m_{\mu}^2 - s)2 -0.25\times \bigg(\frac{cos_w}{sin_w} + \frac{sin_w}{cos_w}\bigg) \bigg(\frac{-y_{\psi}}{\sqrt{2}} \bigg)\bigg(\frac{-y_{\psi}}{\sqrt{2}}\bigg)\bigg( \frac{cos_{w}^{2}+sin_{w}^{2}}{2 cos_{w}sin_{w}}\bigg) \\\nonumber &\times((-2 m_{\psi}^2 + 2 m_{\eta_R}^2 -
2 m_{\psi} m_{\mu} + m_{\mu}^2 - s) \\\nonumber \quad & \times s \sqrt{\frac{1}{
    s^4}(m_{\psi}^4 + (m_{\eta_R}^2 - s)^2 - 
       2 m_{\psi}^2 (m_{\eta_R}^2 + s)) (m_{Z}^4 + (m_{\mu}^2 - s)^2 - 
       2 m_{Z}^2 (m_{\mu}^2 + s))} 
       \\\nonumber &+
(-2 m_{\psi}^4 - 4 m_{\psi}^3 m_{\mu} + m_{\eta_I}^2 (-2 m_{\eta_R}^2 - m_{\mu}^2 + s) + 
  m_{\psi}^2 (2 m_{\eta_I}^2 + 2 m_{\eta_R}^2 - 2 m_{Z}^2 - m_{\mu}^2 + s)   \\\nonumber &+
m_{\mu}^2 (m_{\mu}^2 + s) + 
  2 m_{\psi} m_{\mu} (m_{\eta_I}^2 + m_{\eta_R}^2 - m_{Z}^2 + m_{\mu}^2 + s)) 
          \\\nonumber &
\log\bigg[(m_{\psi}^2 (m_{Z}^2 - m_{\mu}^2 + s) + m_{\eta_R}^2 (-m_{Z}^2 + m_{\mu}^2 + s) + 
   s (-2 m_{\eta_I}^2 + m_{Z}^2 + m_{\mu}^2 - s \\\nonumber \quad &- 
 s \sqrt{\frac{1}{s^4}(m_{\psi}^4 + (m_{\eta_R}^2 - s)^2 - 2 m_{\psi}^2 (m_{\eta_R}^2 + s))
 (m_{Z}^4 + (m_{\mu}^2 - s)^2 - 2 m_{Z}^2 (m_{\mu}^2 + s))}))/  
 \end{align}
 \newpage
 \begin{align}
           \\\nonumber &
  (m_{\psi}^2 (m_{Z}^2 - m_{\mu}^2 + s) + m_{\eta_R}^2 (-m_{Z}^2 + m_{\mu}^2 + s) + 
  s (-2 m_{\eta_I}^2 +  m_{Z}^2 + m_{\mu}^2 - s \\\nonumber \quad & + 
      s \sqrt{\frac{1}{s^4}(m_{\psi}^4 + (m_{\eta_R}^2 - s)^2 - 
            2 m_{\psi}^2 (m_{\eta_R}^2 + s)) (m_{Z}^4 + (m_{\mu}^2 - s)^2 - 
            2 m_{Z}^2 (m_{\mu}^2 + s))}))\bigg]) \\\nonumber \quad & +
            2 \bigg(-2\times \frac{sin_w}{cos_w}\bigg)^2 \bigg(\frac{-y_{\psi}}{\sqrt{2}}\bigg)\bigg(\frac{-y_{\psi}}{\sqrt{2}}\bigg) ((4 (6 m_{\psi}^2 - m_{Z}^2) (m_{\psi}^2 - m_{\eta_R}^2 + 
            2 m_{\psi} m_{\mu} + m_{\mu}^2) s^3 \\\nonumber \quad & \times     \sqrt{\frac{1}{s^4}(m_{\psi}^4 + (m_{\eta_R}^2 - s)^2 - 
               2 m_{\psi}^2 (m_{\eta_R}^2 + s)) (m_{Z}^4 + (m_{\mu}^2 - s)^2 - 
               2 m_{Z}^2 (m_{\mu}^2 + s))})/\\\nonumber \quad & ((m_{\psi}^2 (m_{Z}^2 - m_{\mu}^2 + s) - 
            m_{\eta_R}^2 (m_{Z}^2 - m_{\mu}^2 + s) + 
            s (-m_{Z}^2 - m_{\mu}^2 + s   
           \\\nonumber &+ 
               s \sqrt{\frac{1}{s^4}(m_{\psi}^4 + (m_{\eta_R}^2 - s)^2 - 
                    2 m_{\psi}^2 (m_{\eta_R}^2 + s)) (m_{Z}^4 + (m_{\mu}^2 - s)^2 - 
                    2 m_{Z}^2 (m_{\mu}^2 + s))}))\\\nonumber \quad & \times (m_{\psi}^2 (m_{Z}^2 - m_{\mu}^2 + 
               s) - m_{\eta_R}^2 (m_{Z}^2 - m_{\mu}^2 + s) - 
            s (m_{Z}^2 + m_{\mu}^2 \\\nonumber \quad & + 
               s (-1 +  \sqrt{\frac{1}{s^4}(m_{\psi}^4 + (m_{\eta_R}^2 - s)^2 - 
                    2 m_{\psi}^2 (m_{\eta_R}^2 + s)) (m_{Z}^4 + (m_{\mu}^2 - s)^2 - 
                    2 m_{Z}^2 (m_{\mu}^2 + s))})))) \\\nonumber &- (6 m_{\psi}^2 + 
          6 m_{\psi} m_{\mu} + m_{\mu}^2 - 
          s) \\\nonumber \quad & \times
          \log[\bigg(-m_{\psi}^2 (m_{Z}^2 - m_{\mu}^2 + s) + 
            m_{\eta_R}^2 (m_{Z}^2 - m_{\mu}^2 + s) + 
            s \bigg(m_{Z}^2 + m_{\mu}^2 
         &- 
               s (1 + \sqrt{(1/(
                    s^4))(m_{\psi}^4 + (m_{\eta_R}^2 - s)^2 - 
                    2 m_{\psi}^2 (m_{\eta_R}^2 + s)) (m_{Z}^4 + (m_{\mu}^2 - s)^2 - 
                    2 m_{Z}^2 (m_{\mu}^2 + s))})\bigg)\bigg)/\\\nonumber \quad & (-m_{\psi}^2 (m_{Z}^2 - 
               m_{\mu}^2 + s) + m_{\eta_R}^2 (m_{Z}^2 - m_{\mu}^2 + s) + 
            s (m_{Z}^2 + m_{\mu}^2 \\\nonumber \quad & + 
               s (-1 + \sqrt{(1/(
                    s^4))(m_{\psi}^4 + (m_{\eta_R}^2 - s)^2 - 
                    2 m_{\psi}^2 (m_{\eta_R}^2 + s)) (m_{Z}^4 + (m_{\mu}^2 - s)^2 - 
                    2 m_{Z}^2 (m_{\mu}^2 + s))})))]) \\\nonumber \quad &- \bigg(\frac{1}{m_{\mu}^2 - s}\bigg)
    4 \bigg(-2 \frac{sin_w}{cos_w}\bigg) \bigg(-0.25 \frac{cos_w}{sin_w} + \frac{sin_w}{cos_w}\bigg) \bigg(\frac{-y_{\psi}}{\sqrt{2}}\bigg)\bigg( \frac{-y_{\psi}}{\sqrt{2}} \bigg)\\\nonumber &\times (s (m_{\psi}^2 - m_{\eta_R}^2 + 3 m_{\psi} m_{\mu} + 
          m_{\mu}^2 + s)\\\nonumber & \sqrt{(1/(
          s^4))(m_{\psi}^4 + (m_{\eta_R}^2 - s)^2 - 
             2 m_{\psi}^2 (m_{\eta_R}^2 + s)) (m_{Z}^4 + (m_{\mu}^2 - s)^2 - 
             2 m_{Z}^2 (m_{\mu}^2 + s))} \\\nonumber &- (m_{\psi}^4 + m_{\eta_R}^4 + 6 m_{\psi}^3 m_{\mu} +
           m_{\mu}^2 (-m_{Z}^2 + s) - m_{\eta_R}^2 (m_{\mu}^2 + s) \\\nonumber &+ 
          m_{\psi} m_{\mu} (-6 m_{\eta_R}^2 - 2 m_{Z}^2 + 3 (m_{\mu}^2 + s)) - 
          m_{\psi}^2 (2 m_{\eta_R}^2 + m_{Z}^2 \\\nonumber &- 
             2 (4 m_{\mu}^2 + s))) \log\bigg[(-m_{\psi}^2 (m_{Z}^2 - m_{\mu}^2 + s) + 
            m_{\eta_R}^2 (m_{Z}^2 - m_{\mu}^2 + s) + 
            s (m_{Z}^2 \\\nonumber &+ m_{\mu}^2 - 
               s (1 + \sqrt{\frac{1}
                    {s^4}(m_{\psi}^4 + (m_{\eta_R}^2 - s)^2 - 
                    2 m_{\psi}^2 (m_{\eta_R}^2 + s)) (m_{Z}^4 + (m_{\mu}^2 - s)^2 - 
                    2 m_{Z}^2 (m_{\mu}^2 + s))})))/\\\nonumber \quad & (-m_{\psi}^2 (m_{Z}^2 - 
               m_{\mu}^2 + s) + m_{\eta_R}^2 (m_{Z}^2 - m_{\mu}^2 + s) + 
            s (m_{Z}^2 + m_{\mu}^2 + 
               s \\\nonumber \quad & \times(-1 + \sqrt{\frac{1}
                    {s^4}(m_{\psi}^4 + (m_{\eta_R}^2 - s)^2 - 
                    2 m_{\psi}^2 (m_{\eta_R}^2 + s)) (m_{Z}^4 + (m_{\mu}^2 - s)^2 - 
                    2 m_{Z}^2 (m_{\mu}^2 + 
                    s))})))\bigg]))\bigg]
                    \\\nonumber &
\end{align}

\noindent The general form for the thermally averaged annihilation cross section is 
\begin{eqnarray}
  <  \sigma v >_{AB} & =& \dfrac{1}{2T m_{A}^{2}m_{B}^{2}\kappa_{2}\left(m_{A}/T \right)\kappa_{2}\left( m_{B}/T \right)} \int_{(m_{A}+m_{B})^{2}}^{\infty} \sigma_{AB\longrightarrow CD} \\ && (s-(m_{A}+m_{B})^{2})  \sqrt{s} \kappa_{1} \left(\sqrt{s}/T \right). 
\end{eqnarray}

\noindent Here $\sigma _{AB \longrightarrow CD}$ is the cross section for the process $A+B \longrightarrow C+D$, $\kappa_{i}$s are modified Bessel functions of order $i$ and $T$ is the temperature.  

\section{CALCULATION OF SPHALERON CONVERSION FACTOR  }\label{appen2}
Here we derive the calculation of sphaleron conversion factor in our model according to the standard procedure described in \cite{Harvey:1990qw}. For a relativistic particle X with spin s and degrees of freedom (dof) $g_X$ , the relationship between the particle-antiparticle asymmetry and the particle's chemical potential is expressed as

\begin{equation}
Y_X - Y_{\bar{X}} =\frac{g_X T^2}{6} 
\begin{cases} 
 \mu_X & \text{for fermions} \\
2\mu_X & \text{for bosons}
\end{cases}
\end{equation}
The number of chemical potentials (or asymmetries) corresponds to the number of distinct particle species present in the plasma. However, this number is significantly reduced by constraints arising from chemical equilibrium conditions and conservation laws that govern the early universe as:

\begin{enumerate}
    \item All the gauge bosons have vanishing chemical potential, imposing the equality of chemical potentials among electroweak and color multiplets.
    \item Regardless of the temperature of the universe, the electric charge must be conserved, leading to the following constraint,
    \begin{equation}\label{charge}
        Q = \sum_{i}^{N_{f}} (\mu_{Q_i} + 2 \mu_{u_i} - \mu_{d_i} - \mu_{l_i}-\mu_{e_i} ) + \sum_{i}^{m}2 \mu_{\phi} + \sum_{i}^{n}6\mu_{\psi} = 0.
    \end{equation}
Here, $\mu_{Q_i}, \mu_{u_i}, \mu_{d_i}, \mu_{l_i} and \mu_{e_i}$ represent the chemical potentials for left-handed quark doublets, right-handed up-type quarks, right-handed down-type quarks, left-handed lepton doublets, right-handed charged leptons, respectively. The parameter $N_f$ denotes the number of fermion generations present in the model. Similarly, $\mu_{\phi}$ and $\mu_{\psi_i}$  represent the chemical potential for the scalar doublets and the fermion triplet, respectively, where
m and n indicate the number of scalar doublets and the number of fermion triplet generations in the model.
 \item Nonperturbative electroweak sphaleron and QCD instanton processes, while in thermal equilibrium, imposes the following constraints
 \begin{eqnarray}
     \sum_{i}^{N_f}(3 \mu_{Q_i}+\mu_{l_i}) = 0, \\
     \sum_{i}^{N_f}(2 \mu_{Q_i}-\mu_{d_i}-\mu_{u_i}) = 0
 \end{eqnarray}
 \item All the SM Yukawa interactions and electroweak sphalerons in equilibrium also impose certain conditions as:
 \begin{eqnarray}\nonumber
     \mu_{u_i} - \mu_{Q_i} - \mu_{\phi} = 0, \qquad\phi^0 \longleftrightarrow \bar{u}_L + u_R, \\\nonumber
     \mu_{d_i} - \mu_{Q_i} + \mu_{\phi} = 0, \qquad\phi^0 \longleftrightarrow \bar{d}_R + d_L, \\\nonumber
     \mu_{e_i} - \mu_{l_i} + \mu_{\phi} = 0, \qquad\phi^0 \longleftrightarrow \bar{e}_{iR} + e_{iL},\\
     \mu_{\psi_i} - \mu_{l_i} - \mu_{\phi} = 0, \qquad\psi^0 \longleftrightarrow \phi^{0} + l_{iL}.     
 \end{eqnarray}
\end{enumerate}
Assuming equilibrium among different generations, the generation index $i$ can be dropped from the above equations. Replacing $\mu_{u_i}, \mu_{d_i}, \mu_{e_i}$ and $\mu_{\psi_i}$ in Eq. (\ref{charge}), $\mu_{\phi}$ can be expressed in terms of $\mu_{Q_i} \equiv \mu_{Q}$ as 
\begin{eqnarray}
    \mu_{\phi} = -\frac{8 N_f - 18n}{4 N_f + 2m +6 n}\mu_{Q}.
\end{eqnarray}
The baryon number B in terms of chemical potential is written as 
\begin{equation}\label{baryon}
   B= \sum_{i}^{N_f}(2 \mu_{Q_i} + \mu_{u_i}+\mu_{d_i}) = 4N_{f}\mu_{Q}.
\end{equation}
Similarly, the lepton number for muon is written as:
\begin{eqnarray}\label{lepton}
    L_{\mu} = (2 \mu_{l_\mu} + \mu_{\mu_R})  = -9  \mu_{Q}+ \bigg[\frac{(8 N_f-18n)}{(4 N_f + 2m+6n)}\bigg]\mu_{Q}.
\end{eqnarray}
From Eq. (\ref{baryon}) and Eq. (\ref{lepton}), we can write 
\begin{eqnarray}\nonumber
B &=& -\frac{16N_f^{2} + 8mN_{f}+24nN_{f}}{28N_{f}+18m+72n} L_{\mu}, \\
    B &=& \frac{16N_{f}^{2}+8mN_{f}+24nN_{f}}{16N_{f}^{2}+18mN_{f}+24nN_{f}+28N_{f}+18m+72n}B-L_{\mu}.
\end{eqnarray}
For our model, with $N_f = 3$, $m = 2$ and $n=1$,  $ B = \frac{33}{57}(B-L_{\mu})$.
\typeout{}
\bibliographystyle{unsrt}
\bibliography{ref}

\end{document}